\documentclass[twocolumn,showpacs,preprintnumbers,amsmath,amssymb,superscriptaddress,aps,prc]{revtex4-1}

\usepackage{rotating}
\usepackage{graphicx}
\usepackage{dcolumn}
\usepackage{bm}
\usepackage{mathtools}
\usepackage{amsmath}
\usepackage{textcomp}
\usepackage{longtable}
\usepackage{color}
\usepackage{hyperref}
\hypersetup{
 colorlinks,
 linkcolor={blue},
 citecolor={blue},
 urlcolor={blue},
 breaklinks=true
}

\begin{document}

\title{Prompt-delayed $\gamma$-ray spectroscopy of neutron-rich $^{119,121}$In isotopes}

\newcommand{\IPHC}{Universit\'e de Strasbourg, CNRS, IPHC UMR 7178, F-67000 Strasbourg, France}
\newcommand{\IPNL}{Univ. Lyon, Universit\'e Lyon 1, CNRS/IN2P3, IPN-Lyon, F-69622, Villeurbanne, France }
\newcommand{\GANIL}{GANIL, CEA/DRF-CNRS/IN2P3, Bd Henri Becquerel, BP 55027, F-14076 Caen Cedex 05, France}
\newcommand{\CSNSM}{CSNSM, Univ. Paris-Sud, CNRS/IN2P3, Universit\'e Paris-Saclay, 91405 Orsay, France}
\newcommand{\IJCLAB}{Universit\'e Paris-Saclay, CNRS/IN2P3, IJCLab, 91405  Orsay, France}
\newcommand{\TUDarmstadt}{Institut f\"ur Kernphysik, Technische Universit\"at Darmstadt, D-64289 Darmstadt, Germany}
\newcommand{\IFIC}{Instituto de F\'isica Corpuscular, CSIC-Universitat de Val\`encia, E-46980 Val\`encia, Spain}
\newcommand{\Canada}{Department of Chemistry, Simon Fraser University, Burnaby, British Columbia, Canada}
\newcommand{\Legnaro}{INFN, Laboratori Nazionali di Legnaro, Via Romea 4, I-35020 Legnaro, Italy}
\newcommand{\IPN}{Institut de Physique Nucl\'eaire, IN2P3-CNRS, Univ. Paris Sud, Universit\'e Paris Saclay, 91406 Orsay Cedex, France}
\newcommand{\Debrecen}{Institute for Nuclear Research of the Hungarian Academy of Sciences, Pf.51, H-4001, Debrecen, Hungary}
\newcommand{\Somerset}{iThemba LABS, National Research Foundation, P.O.Box 722, Somerset West,7129 South Africa}
\newcommand{\Padova}{INFN Sezione di Padova, I-35131 Padova, Italy}
\newcommand{\UPadova}{Dipartimento di Fisica e Astronomia dell'Universit\`a di Padova, I-35131 Padova, Italy}
\newcommand{\GSI}{GSI, Helmholtzzentrum f\"ur Schwerionenforschung GmbH, D-64291 Darmstadt, Germany}
\newcommand{\Milano}{INFN, Sezione di Milano, I-20133 Milano, Italy}
\newcommand{\LPSC}{LPSC, Universit\'e Grenoble-Alpes, CNRS/IN2P3, 38026 Grenoble Cedex, France}
\newcommand{\IRFU}{IRFU, CEA/DRF, Centre CEA de Saclay, F-91191 Gif-sur-Yvette Cedex, France}
\newcommand{\UMilano}{Dipartimento di Fisica, Universit\`a di Milano, I-20133 Milano, Italy}
\newcommand{\STFC}{STFC Daresbury Laboratory, Daresbury, Warrington, WA4 4AD, UK}
\newcommand{\LBNL}{Nuclear Science Division, Lawrence Berkeley National Laboratory, Berkeley, California 94720, USA}
\newcommand{\IFJ}{Institute of Nuclear Physics PAN, 31-342 Krak\'ow, Poland}
\newcommand{\VECC}{Variable Energy Cyclotron Centre, 1/AF Bidhan Nagar, Kolkata 700064, India}
\newcommand{\TIFR}{Department of Nuclear and Atomic Physics, Tata Institute of Fundamental Research,Mumbai, 400005, India}
\newcommand{\ILL}{Institut Laue-Langevin, F-38042 Grenoble Cedex, France}
\newcommand{\HBNI}{Homi Bhabha National Institute, Training School Complex, Anushaktinagar, Mumbai-400094, India}

\author{S. Biswas}
\affiliation{\GANIL}

\author{A.~Lemasson}
\email{lemasson@ganil.fr}
\affiliation{\GANIL}

\author{M.~Rejmund}
\affiliation{\GANIL}

\author{A.~Navin}
\affiliation{\GANIL}

\author{Y.H.~Kim}
\altaffiliation[Present address: ]{\ILL}
\affiliation{\GANIL}

\author{C.~Michelagnoli}
\altaffiliation[Present address: ]{\ILL}
\affiliation{\GANIL}

\author{I.~Stefan}
\affiliation{\IJCLAB}

\author{R.~Banik}
\affiliation{\VECC}
\affiliation{\HBNI}

\author{P. Bednarczyk}
\affiliation{\IFJ}

\author{Soumik~Bhattacharya}
\affiliation{\VECC}
\affiliation{\HBNI}

\author{S.~Bhattacharyya}
\affiliation{\VECC}
\affiliation{\HBNI}

\author{E.~Cl\'{e}ment}
\affiliation{\GANIL}

\author{H. L. Crawford}
\affiliation{\LBNL}

\author{G.~de~France}
\affiliation{\GANIL}

\author{P. Fallon}
\affiliation{\LBNL}

\author{G.~Fr\'{e}mont}
\affiliation{\GANIL}

\author{J.~Goupil}
\affiliation{\GANIL}

\author{B.~Jacquot}
\affiliation{\GANIL}

\author{H.J.~Li}
\affiliation{\GANIL}

\author{J.~Ljungvall}
\affiliation{\IJCLAB}


\author{A. Maj}
\affiliation{\IFJ}

\author{L.~M\'enager}
\affiliation{\GANIL}

\author{V.~Morel}
\affiliation{\GANIL}

\author{R.~Palit}
\affiliation{\TIFR}

\author{R.M.~P\'erez-Vidal}
\affiliation{\IFIC}

\author{J.~Ropert}
\affiliation{\GANIL}



\date{\today}

\begin{abstract}

{\bf Background:} The $Z = 50$ shell closure, near $N=82$, is unique in the sense that it is the only shell closure with the spin-orbit partner orbitals, $\pi g_{9/2}$ and $\pi g_{7/2}$, enclosing the magic gap. The interaction of the proton hole/particle in the above mentioned orbitals with neutrons in the $\nu h_{11/2}$ orbital is an important prerequisite to the understanding of the nuclear structure near $N = 82$ and the $\nu\pi$ interaction.

{\bf Purpose:} To explore the structural similarity between the high-spin isomeric states in In ($Z=49$), Sn ($Z=50$) and Sb ($Z=51$) isotopes from a microscopic point of view. In addition, to understand the role of a proton hole or particle in the spin-orbit partner orbitals, $\pi g_{9/2}$ and $\pi g_{7/2}$, respectively, with neutron holes in the $\nu h_{11/2}$ orbital on these aforementioned isomers.

{\bf Methods:} The fusion and transfer induced fission reaction $^{9}$Be($^{238}$U,~f) with 6.2 MeV/u beam energy, using a unique setup consisting of AGATA, VAMOS++ and EXOGAM detectors, was used to populate through the fission process and study the neutron-rich $^{119,121}$In isotopes. This setup enabled the prompt-delayed $\gamma$-ray spectroscopy of isotopes in the time range of $100~\rm{ns} - 200~\mu\rm{s}$.

{\bf Results:} In the odd-$A$ $^{119,121}$In isotopes, indications of a short half-life $19/2^{-}$ isomeric state, in addition to the previously known $25/2^{+}$ isomeric state, were observed from the present data. Further, new prompt transitions above the $25/2^{+}$ isomer in $^{121}$In were identified along with reevaluation of its half-life.

{\bf Conclusions:} The experimental data were compared with the theoretical results obtained in the framework of large-scale shell-model calculations in a restricted model space. The $\langle \pi g_{9/2} \nu h_{11/2};I \arrowvert \hat{\mathcal{H}}\arrowvert \pi g_{9/2} \nu h_{11/2};I\rangle$ two-body matrix elements of residual interaction were modified to explain the excitation energies and the $B(E2)$ transition probabilities in the neutron-rich In isotopes. The (i) decreasing trend of $E(29/2^{+}) - E(25/2^{+})$ in odd-In (with dominant configuration $\pi g_{9/2}^{-1}\nu h_{11/2}^{-2}$ and maximum aligned spin of $29/2^{+}$) and (ii) increasing trend of $E(27/2^{+}) - E(23/2^{+})$ in odd-Sb (with dominant configuration $\pi g_{7/2}^{+1}\nu h_{11/2}^{-2}$ and maximum aligned spin of $27/2^{+}$) with increasing neutron number could be understood as a consequence of hole-hole and particle-hole interactions, respectively.

\end{abstract}


\maketitle

 
\section{\label{sec:Intro}Introduction}

The protons and neutrons in an atomic nucleus are held together by the dominant strong nuclear force. However, due to the limited theoretical knowledge on the strong force (governed by QCD), the structure of the nucleus is being described using a variety of models. Out of these, the most successful models are the nuclear shell model by Maria Goeppert Mayer and Jensen, and the collective model by Bohr and Mottelson. The empirical one-body nuclear spin-orbit force, introduced by Mayer and Jensen, was able to successfully explain all the experimentally observed magic numbers~\cite{ma49, ha49}. The spin-orbit term led to the splitting of the spin-orbit partner orbitals ($j_{>}=l+s$ and $j_{<}=l-s$), with one of the partners ($j_{>}$) generally moved into the lower oscillator shell, and called the intruder orbital. Such examples of intruder orbitals are the $f_{7/2}$ (leading to the magic number $28$), $g_{9/2}$ (leading to magic number $50$), $h_{11/2}$ (leading to the magic number $82$), and $i_{13/2}$ (leading to the magic number $126$). The magic number $50$ is unique as the nuclei around this shell closure have holes and particles occupying adjacent spin-orbit partners, $j_{<}=g_{7/2}$ and $j_{>}=g_{9/2}$.

In neutron-rich In ($Z = 49$) and Sb ($Z = 51$) isotopes, the valence proton hole and particle occupy the adjacent spin-orbit partners $g_{9/2}$ and $g_{7/2}$, respectively, while the valence neutron holes dominantly occupy the $h_{11/2}$ orbital. Even-$A$ $^{118-130}$Sn isotopes possess $7^{-}$ and $10^{+}$ isomers, with dominant neutron $\nu h_{11/2}^{-1}d_{3/2}^{-1}$ and $\nu h_{11/2}^{-2}$ configurations, respectively~\cite{fo81, pi11, as12, is14}. Odd-$A$ $^{121-131}$Sb isotopes have $19/2^{-}$ and $23/2^{+}$ isomers with dominant $\pi g_{7/2} \nu h_{11/2}^{-1}d_{3/2}^{-1}$ and $\pi g_{7/2} \nu h_{11/2}^{-2}$ configurations~\cite{ju07, wa09, wa09epja, ge03, ge00}, with an additional proton particle in $g_{7/2}$ coupled to the $7^{-}$ and $10^{+}$ isomers in even-$A$ Sn isotopes, respectively. It should be noted that the $23/2^+$ isomers in Sb isotopes do not correspond to maximally aligned configurations ($I^{\pi}_{Max} = 27/2^{+}$) but rather to $I^{\pi}_{Max}-2 = 23/2^+$, illustrating the influence of spin and seniority mixing in these isotopes. From this analogy, the odd-$A$ $^{119-129}$In isotopes would be expected to have $21/2^{-}$ and $25/2^{+}$ isomers with dominant $\pi g_{9/2}^{-1} \nu h_{11/2}^{-1}d_{3/2}^{-1}$ and $\pi g_{9/2}^{-1} \nu h_{11/2}^{-2}$ configurations, having an additional proton hole in $g_{9/2}$ coupled to the $7^{-}$ and $10^{+}$ isomers in even-$A$ Sn, respectively.

\begin{center}
\begin{table}[b]
\caption{\label{tab:litt} Isomeric states half-lives and tentative assignments reported in the literature for odd-$A$ In isotopes $^{119,129}$In.}
 \begin{tabular}{p{0.24\columnwidth} p{0.24\columnwidth} p{0.24\columnwidth} p{0.21\columnwidth}} 
 \hline
  Nucleus & $J^\pi$ & T$_{1/2}$  & Reference \\
 \hline
 \hline
 $^{119}$In & ($25/2^+$) & 240~(25)~ns & \cite{lu02} \\ 
 \hline
 $^{121}$In & ($25/2^+$) & 350~(50)~ns  & \cite{lu02} \\ 
  & ($21/2^-$) & 17~(2)~$\mu$s &\cite{re10} \\ 
 \hline
 $^{123}$In & ($17/2^-$) &  1.4~(2)~$\mu$s & \cite{re10,sc04} \\ 
  & ($21/2^-$) & $>100$~$\mu$s &\cite{re10} \\ 
 \hline
 $^{125}$In & ($25/2^+$) &  5.0~(15)~ms & \cite{sc04} \\ 
  & ($21/2^-$) & 9.4~(6)~$\mu$s &\cite{re10,sc04} \\ 
 \hline
 $^{127}$In & ($29/2^+$) &  9.0~(2)~$\mu$s & \cite{sc04} \\ 
  & ($21/2^-$) & 1.04~(10)~s &\cite{ba18} \\ 
 \hline
 $^{129}$In & ($29/2^+$) & 110.0~(15)~ms & \cite{sc04,ga04} \\ 
  & ($17/2^-$) & 8.5~(5)~$\mu$s & \cite{sc04} \\ 
  & ($17/2^-$) & 8.7~(7)~$\mu$s & \cite{ga04} \\ 
  & ($17/2^-$) & 11.3~$^{+22}_{-16} \mu$s &\cite{ka12} \\ 
  & ($23/2^-$) & 700~ms &\cite{sc04} \\ 
  & ($23/2^-$) & 670~(10)~ms &\cite{ga04} \\ 
 \hline
\end{tabular}
\end{table}
\end{center}

In the odd-$A$ $^{119-129}$In isotopes, several isomeric states associated to the above mentioned configurations have been reported in Refs.~\cite{lu02,re10,sc04,ba18,ka12,ga04} and are summarized in Table~\ref{tab:litt}. In $^{119}$In, a ($25/2^{+}$) isomeric state was observed with a half-life of $240(25)$~ns but there was no indication of negative-parity isomer~\cite{lu02}. In $^{121}$In, an isomeric state, decaying by the $99$~keV transition and decaying through a $\gamma$ cascade involving 99-214-953-1181~keV was reported in Ref.~\cite{lu02} and Ref.~\cite{re10} with two  different half-lives (350~(50)~ns and 17~(2)~$\mu$s respectively) and tentative assignments ($25/2^{+}$ and $21/2^{-}$ respectively).  In heavier  $^{123-129}$In isotopes, several positive parity  ($23/2^+$, $25/2^+$, $29/2^+$) and negative parity ($17/2^-$ and $21/2^-$) isomeric states were reported in Refs.~\cite{re10,sc04,ba18,ka12,ga04}. These experimental data indicate the need to perform prompt-delayed $\gamma$-ray spectroscopy of the neutron-rich In isotopes for the reliable assignment of states.

In the present manuscript, new high-spin prompt $\gamma$-ray transitions above the ($25/2^{+}$) isomeric state are reported in the odd-$A$ $^{121}$In isotope along with new determination of the lifetime of the isomeric ($25/2^{+}$) state. Further, the existence of new ($19/2^{-}$) isomers in $^{119, 121}$In isotopes is also discussed. These new experimental results are discussed using large-scale shell-model calculations.

\section{\label{sec:Exp}Experimental Details}

A $^{238}$U beam at an energy of 6.2 MeV/u impinging on a $^{9}$Be target was used to populate the neutron-rich fission fragments $^{119,121}$In via fusion- and transfer-induced fission reactions at GANIL. Two targets were used during the experiment : the $1.6$~$\mu$m and $5$~$\mu$m targets were bombarded for 32 hours and 160 hours respectively with typical beam intensities of  1~pnA and 0.45~pnA respectively.  The two datasets were combined for the present data analysis. The experimental setup consisted of a combination of AGATA $\gamma$-ray tracking array~\cite{cl16,ak12}, VAMOS++ spectrometer~\cite{re11} and the EXOGAM array~\cite{si00}. The AGATA array was placed at $13.5$~cm from the target position, which consisted of $32$ crystals~\cite{cl16}. The focal plane detection system of the VAMOS++ spectrometer, placed at $20^{\circ}$ relative to the beam axis, consisted of a Multi-Wire Proportional Counter (MWPC), two drift chambers and a segmented ionization chamber. It was used for the unambiguous isotopic identification of the fission fragments ($Z$, $A$, $q$)~\cite{re11, na14, nare, ki17}. The velocity vector of the recoiling ions (measured by Dual Position-Sensitive MWPC (DPS-MWPC) detector~\cite{va16}), placed at the entrance of the VAMOS++ spectrometer, and the $\gamma$-ray emission angle (determined using AGATA) were used to obtain the energy of the Doppler corrected prompt $\gamma$ rays ($\gamma_{P}$), on an event-by-event basis. The delayed $\gamma$ rays ($\gamma_{D}$) were detected using seven EXOGAM HPGe Clover detectors~\cite{si00}, arranged in a wall like configuration at the focal plane of VAMOS++, behind the ionization chamber. The decays curves were obtained from the time difference ($t_{decay}$) between the signals from the DPS-MWPC detector placed at the entrance of VAMOS++ and the EXOGAM HPGe detectors placed at the focal plane, and thus the time measured is independent of the time of flight.  Due to the logic delays used during the experiment the presented time ($t_{decay}$) has $800$~ns offset, the true reaction time is located therefore at $t_{decay} = 800$~ns. Additional information on the background subtraction, efficiency evaluation, half-life evaluation are given in the Refs.~\cite{ki17, sb19}. The uncertainties in the energy of prompt and delayed $\gamma$ ray transitions is $\sim$ 1~keV. The efficiencies for the prompt and delayed $\gamma$ rays were determined separately. The spin-parities were assigned based on systematics and shell model calculations. The tabulated values of the energies and relative intensities of the prompt and delayed $\gamma$-ray transitions are provided as Supplemental Material~\cite{Sup}. 

\section{\label{sec:Res}Experimental Results}

\subsection{\label{sec:119in}$^{119}$In}

\begin{figure}[]
\includegraphics[width=1.0\columnwidth]{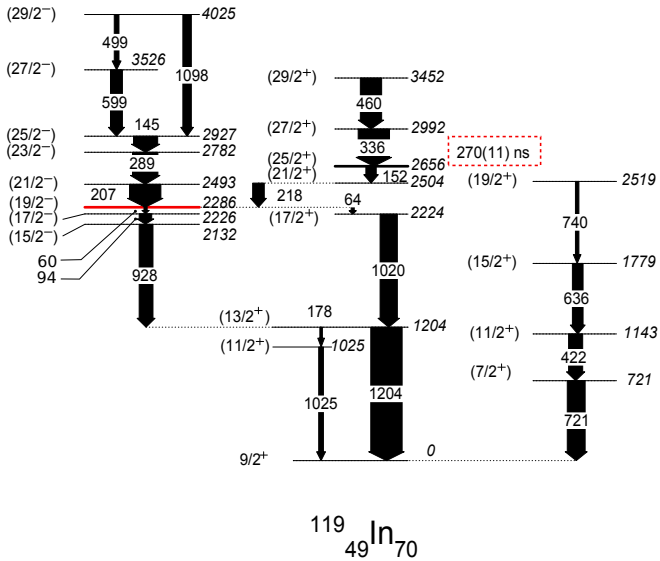}
\caption{\label{fig:119in_fig1} (Color online) The level scheme of $^{119}$In. The widths of the arrows represent the intensities of the transitions.  The newly identified and the previously known isomeric states are indicated by a thick red and black line, respectively. The previously known half-life but remeasured in this work and consistent with the previous work is marked by a red dashed box.}
\end{figure}

\begin{figure}[]
\includegraphics[width=1.0\columnwidth]{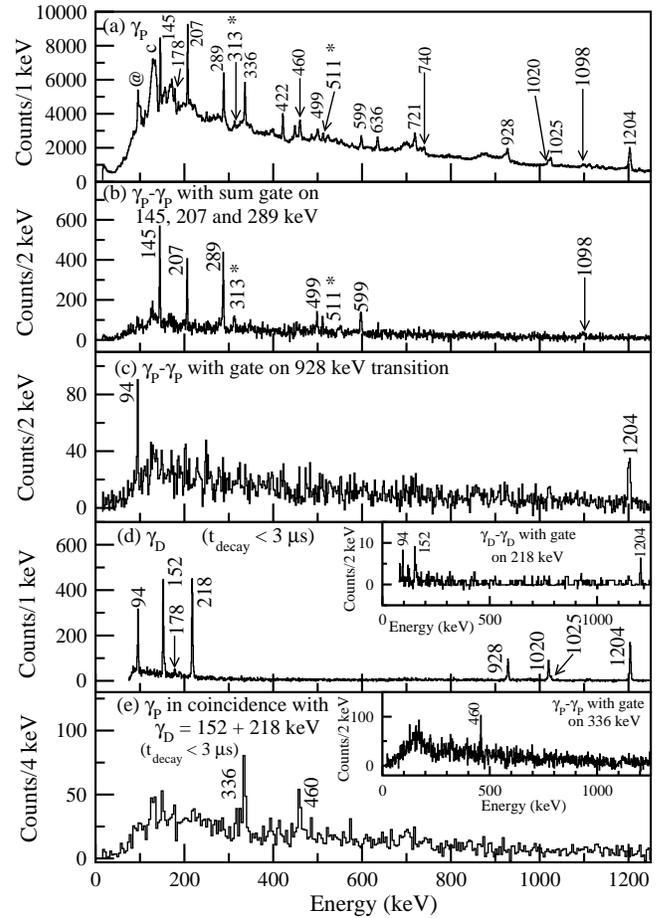}
\caption{\label{fig:119in_fig2} $A-$ and $Z-$gated $\gamma$-ray spectrum for $^{119}$In: (a) The tracked Doppler corrected prompt singles $\gamma$-ray ($\gamma_{P}$) spectrum with the new $\gamma$-ray transitions marked with asterisk. (b) Tracked Doppler corrected prompt $\gamma_{P}$-$\gamma_{P}$ coincidence spectrum with sum gate on the $145$, $207$, and $289$~keV $\gamma$-ray transitions. (c) Tracked Doppler corrected prompt $\gamma_{P}$-$\gamma_{P}$ coincidence spectrum with gate on the $928$~keV $\gamma$-ray. (d) The delayed singles $\gamma$-ray ($\gamma_{D}$) spectra for $t_{decay}<3~\mu$s. (e) Tracked Doppler corrected $\gamma_{P}$ in coincidence with $\gamma_{D}$ = $152$ and $218$~keV $\gamma$ rays (for $t_{decay}<3~\mu$s). The inset shows the tracked Doppler corrected prompt $\gamma_{P}$-$\gamma_{P}$ coincidence spectrum with gate on the $336$~keV $\gamma$-ray.}
\end{figure}

\begin{figure}[h]
\includegraphics[width=1.0\columnwidth]{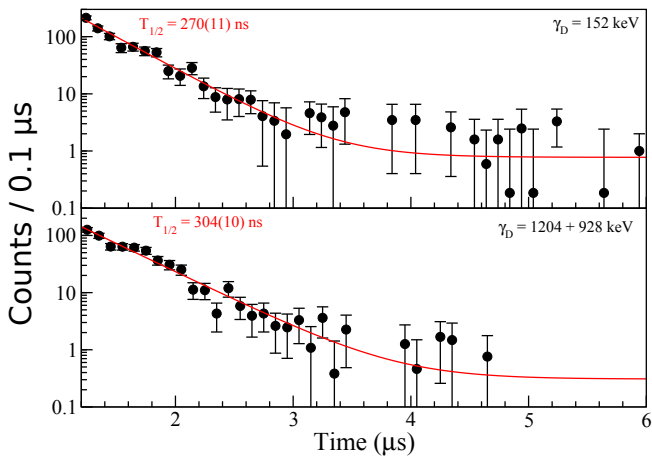}
\caption{\label{fig:119in_fig3} Decay curves along with the fits for the different transitions in $^{119}$In: (a) $152$~keV and (b) sum of $1204$ and $928$~keV transitions. }
\end{figure}

The level scheme for the $^{119}$In isotope, obtained from the present work is shown in Fig.~\ref{fig:119in_fig1}.  Previous measurement on the high-spin states in $^{119}$In was reported in Ref.~\cite{lu02}. The $A$- and $Z$-gated tracked Doppler corrected prompt singles $\gamma$-ray ($\gamma_{P}$) spectrum, for the $1.6\times10^6$ $^{119}$In ions identified in VAMOS++ in coincidence with prompt $\gamma$-ray, is shown in Fig.~\ref{fig:119in_fig2}(a).  The prompt $\gamma$-ray transition emitted by the complementary fission fragments~\cite{na14, nare} (mainly from the fusion-fission process) are observed and identified (marked by "c") in the low energy part of the $\gamma$-ray spectra. The random coincidences with X-ray emitted by $^{238}$U is also observed and is marked by an $@$ symbol. All the previously known $\gamma$ rays were observed in the present spectrum, except for the $60$~keV ($19/2^{-} \rightarrow 17/2^{-}$) and $64$~keV ($19/2^{-} \rightarrow 17/2^{+}$) $\gamma$-ray transitions suggesting a lifetime of the ($19/2^{-}$) state. In addition, two new prompt $\gamma$ rays, $313$ and $511$~keV are observed and are marked with an asterisk. The tracked Doppler corrected prompt $\gamma_{P}$-$\gamma_{P}$ coincidence spectrum with a sum gate on the $145$, $207$, and $289$~keV $\gamma$-ray transitions is shown in Fig.~\ref{fig:119in_fig2}(b). The newly observed $313$ and $511$~keV transitions along with the previously observed $145$, $207$, $289$, $499$, $599$ and $1098$~keV transitions are marked in this figure. However, due to lack of enough statistics in the coincidence spectrum, the 313 and 511~keV transitions were not placed in the level scheme in Fig.~\ref{fig:119in_fig1}. The $94$, $928$ and $1204$~keV transitions are not observed, pointing to the fact that the ($19/2^{-}$) state could be an isomeric state. This was confirmed by the observation of only $94$ and $1204$~keV transitions with gate on $928$ keV transition in the prompt $\gamma_{P}$ spectrum, as shown in Fig.~\ref{fig:119in_fig2}(c). The tracked Doppler corrected prompt $\gamma_{P}$-$\gamma_{P}$ coincidence spectrum with a gate on the $721$~keV $\gamma$-ray transition, yielding $422$, $636$ and $740$~keV transitions (not shown in this figure). The delayed $\gamma$-ray spectrum ($\gamma_{D}$) for $t_{decay} < 3~\mu$s is shown in Fig.~\ref{fig:119in_fig2}(d), yielding $94$, $152$, $178$, $218$, $928$, $1020$, $1025$ and $1204$~keV transitions. The 60 and 64~keV transitions are below the detection threshold of the present delayed-$\gamma$-ray setup and cannot be observed. The inset of Fig.~\ref{fig:119in_fig2}(d) shows the delayed $\gamma_{D}$-$\gamma_{D}$ coincidence spectrum with gate on the $218$~keV $\gamma$-ray transition, yielding $94$, $152$ and $1204$~keV transitions. The tracked Doppler corrected $\gamma_{P}$ in coincidence with any $\gamma_{D}$ (for $t_{decay} < 3~\mu$s) is shown in Fig.~\ref{fig:119in_fig2}(e). This spectrum yields the previously identified prompt $\gamma$-ray transitions, $336$ and $460$~keV transitions. The tracked Doppler corrected prompt $\gamma_{P}$-$\gamma_{P}$ coincidence spectrum with a gate on the $336$~keV $\gamma$-ray also yields $460$~keV in coincidence, as shown in the inset of Fig.~\ref{fig:119in_fig2}(e). This shows that the $336$ and $460$~keV transitions are indeed in coincidence and lie above the ($25/2^{+}$) isomer.

Figure~\ref{fig:119in_fig3} shows the half-life fit (one-component) for the decay spectrum upon gating on $152$~keV transitions yielding a value of $T_{1/2} = 270(11)$~ns for the ($25/2^{+}$) state (in agreement with the value of $240 (25)$~ns reported in Ref.~\cite{lu02}). The presently reported half-life corresponds to $B(E2; 25/2^{+} \rightarrow 21/2^{+}) = 19(1)$~e$^{2}$fm$^{4}$. Even with a very short time gate, it was not possible to obtain a $\gamma_{D}$ spectra containing only the $94$, $928$, and $1204$~keV transitions, indicating that the ($19/2^{-}$) state has a short half-life, which is below the sensitivity of the present setup. In order to have an estimation of the half-life of the ($19/2^{-}$) state, the half-life fit (one-component) for the decay spectrum upon gating on the $1204$~keV transition was determined to be $T_{1/2} = 304(10)$~ns. This shows that the half-life of the ($19/2^{-}$) state should be around $3 \rm{ns} < T_{1/2} < 10$~ns, which is compatible with the half-life estimate for an $M1$ $60$~keV or $E1$ $64$~keV transition that is consistent with the non-observation of these transitions in the prompt $\gamma$-ray spectrum.

\subsection{\label{sec:121in}$^{121}$In}

\begin{figure}[]
\includegraphics[width=1.0\columnwidth]{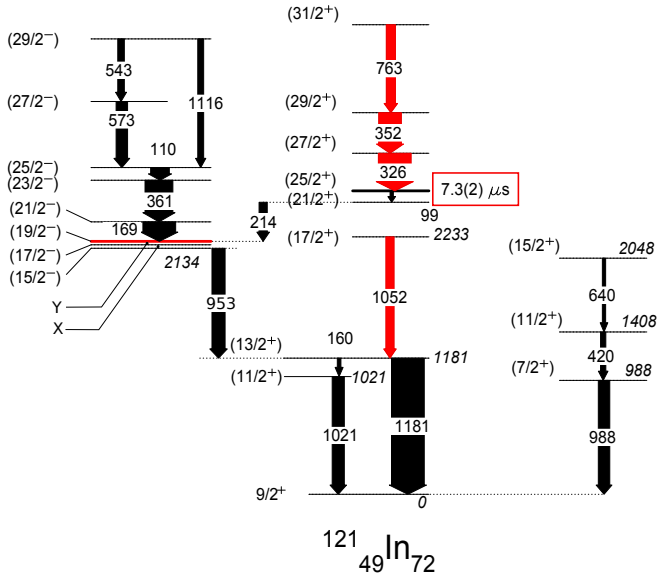}
\caption{\label{fig:121in_fig1} (Color online) The level scheme of $^{121}$In. The widths of the arrows represent the intensities of the transitions.  The newly identified transitions are shown in red. The newly identified and the previously known isomeric states are indicated by a thick red and black line, respectively. The newly redetermined half-life is marked by a red box. The X and Y energies  refers respectively to the energies of the unobserved $(17/2^{-}) \rightarrow (15/2^{-}) $  and  $(19/2^{-}) \rightarrow (17/2^{-}) $ transitions as reported in Ref.~\cite{lu02}. Because of these unobserved transitions, excitation energies of levels above the  $15/2^{-}$ level are not determined. }
\end{figure}

\begin{figure}[]
\includegraphics[width=1.0\columnwidth]{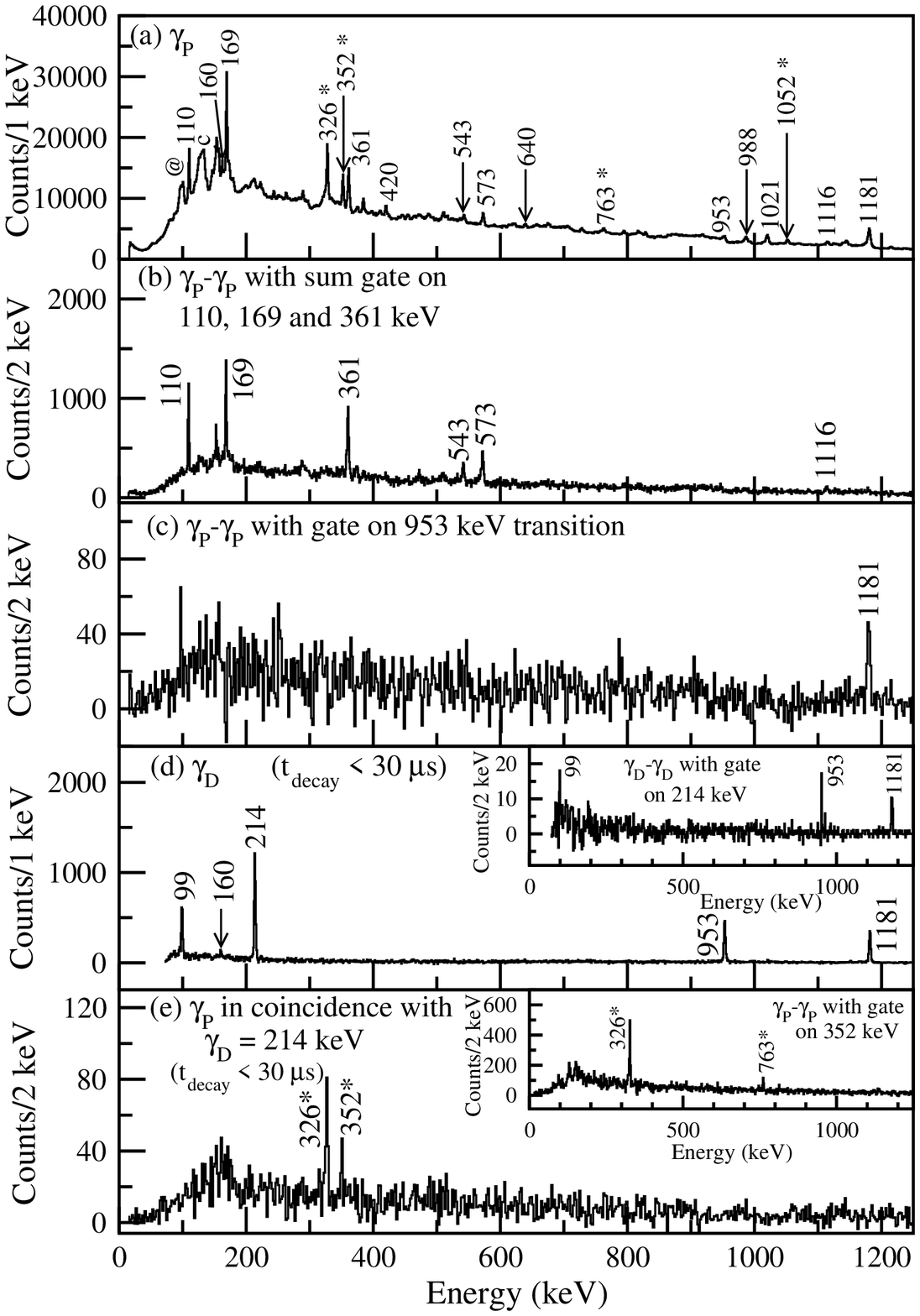}
\caption{\label{fig:121in_fig2} (Color online) $A$- and $Z$-gated $\gamma$-ray spectra for $^{121}$In: (a) The tracked Doppler corrected prompt singles $\gamma$-ray ($\gamma_{P}$) spectrum with the new $\gamma$-ray transitions marked with asterisk. (b) Tracked Doppler corrected prompt $\gamma_{P}$-$\gamma_{P}$ coincidence spectrum with sum gate on the $110$, $169$, and $361$~keV $\gamma$-ray transitions. (c) Tracked Doppler corrected prompt $\gamma_{P}$-$\gamma_{P}$ coincidence spectrum with gate on the $953$~keV $\gamma$-ray. (d) The delayed singles $\gamma$-ray ($\gamma_{D}$) spectra for $t_{decay}<30~\mu$s. The inset in (d) shows the decay curve along with the fit for the $99$~keV transition. (e) Tracked Doppler corrected $\gamma_{P}$ in coincidence with $\gamma_{D} = 214$~keV $\gamma$-ray (for $t_{decay}<30~\mu$s). The inset shows the tracked Doppler corrected prompt $\gamma_{P}$-$\gamma_{P}$ coincidence spectrum with gate on the $352$~keV $\gamma$-ray.}
\end{figure}

The level scheme for the $^{121}$In isotope, obtained from the present work is shown in Fig.~\ref{fig:121in_fig1}. 
The $A$- and $Z$- gated tracked Doppler corrected prompt singles $\gamma$-ray ($\gamma_{P}$) spectrum, for the $4.2\times10^6$ $^{121}$In ions identified in VAMOS++ in coincidence with prompt $\gamma$-ray, is shown in Fig.~\ref{fig:121in_fig2}(a). The prompt $\gamma$-ray transition emitted by the complementary fission fragment~\cite{na14, nare} (mainly from the fusion-fission) are observed and identified (marked by "c") in the low energy part of the $\gamma$-ray spectra. The random coincidences with X-ray emitted by $^{238}$U is also observed and is marked by an $@$ symbol. All the previously known $\gamma$ rays were observed in the present spectrum. In addition, four new prompt transitions $326$, $352$, $763$, and $1052$~keV transitions were observed and these are marked with an asterisk in Fig.~\ref{fig:121in_fig2}(a) and denoted by red in the level scheme (Fig.~\ref{fig:121in_fig1}). The tracked Doppler corrected prompt $\gamma_{P}$-$\gamma_{P}$ coincidence spectrum with sum gate on the $110$, $169$, and $361$~keV $\gamma$-ray transitions is shown in Fig.~\ref{fig:121in_fig2}(b). All the previously observed $110$, $169$, $361$, $543$, $573$ and $1116$~keV transitions are labeled. The $95$3 and $1181$~keV transitions are not observed, pointing to the fact that the ($19/2^{-}$) state is likely to be an isomeric state with a short half-life, which is below the sensitivity of the present setup. This was confirmed by the observation of $1181$~keV transition only with the gate on $953$~keV transition in the prompt $\gamma_{P}$ spectrum, as shown in Fig.~\ref{fig:121in_fig2}(c). The tracked Doppler corrected prompt $\gamma_{P}$-$\gamma_{P}$ coincidence spectrum with a gate on the $988$~keV $\gamma$-ray transition, yielding $420$~keV transition (not shown in this figure). The delayed $\gamma$-ray spectrum ($\gamma_{D}$) for $t_{decay} < 30~\mu$s is shown in Fig.~\ref{fig:121in_fig2}(d), yielding $99$, $160$, $214$, $953$, and $1181$~keV transitions. The inset of Fig.~\ref{fig:121in_fig2}(d) shows the delayed $\gamma_{D}$-$\gamma_{D}$ coincidence spectrum with a gate on the $214$~keV $\gamma$-ray transition, yielding $99$, $953$ and $1181$~keV transitions. The tracked Doppler corrected $\gamma_{P}$ in coincidence with any $\gamma_{D}$ (for $t_{decay} < 30~\mu$s) is shown in Fig.~\ref{fig:121in_fig2}(e). This spectrum yields the newly identified prompt $\gamma$-ray transitions, $326$ and $352$~keV transitions. The tracked Doppler corrected prompt $\gamma_{P}$-$\gamma_{P}$ coincidence spectrum with the sum gate on the newly observed $352$~keV $\gamma$-ray also yields $326$ and $763$ keV in coincidence, as shown in the inset of Fig.~\ref{fig:121in_fig2}(e). This proves that the $326$, $352$ and $763$~keV transitions are indeed in coincidence and lie above the ($25/2^{+}$) isomer.

\begin{figure}[]
\includegraphics[width=1.0\columnwidth]{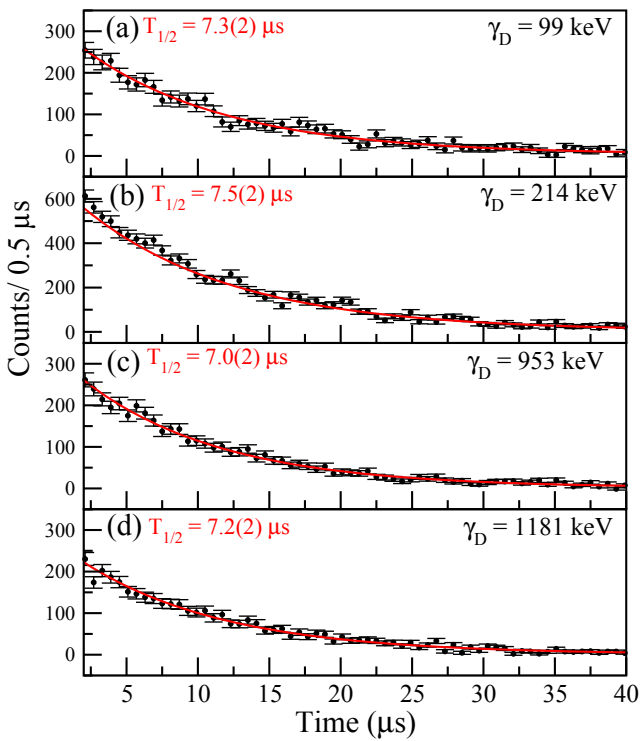}
\caption{\label{fig:121in_fig3} Decay curves along with the fits for the different transitions in $^{121}$In: (a) $99$~keV, (b) $214$~keV, (c) $953$~keV, and (d) $1181$~keV transitions. }
\end{figure}

\begin{figure*}[]
\includegraphics[width=2.1\columnwidth]{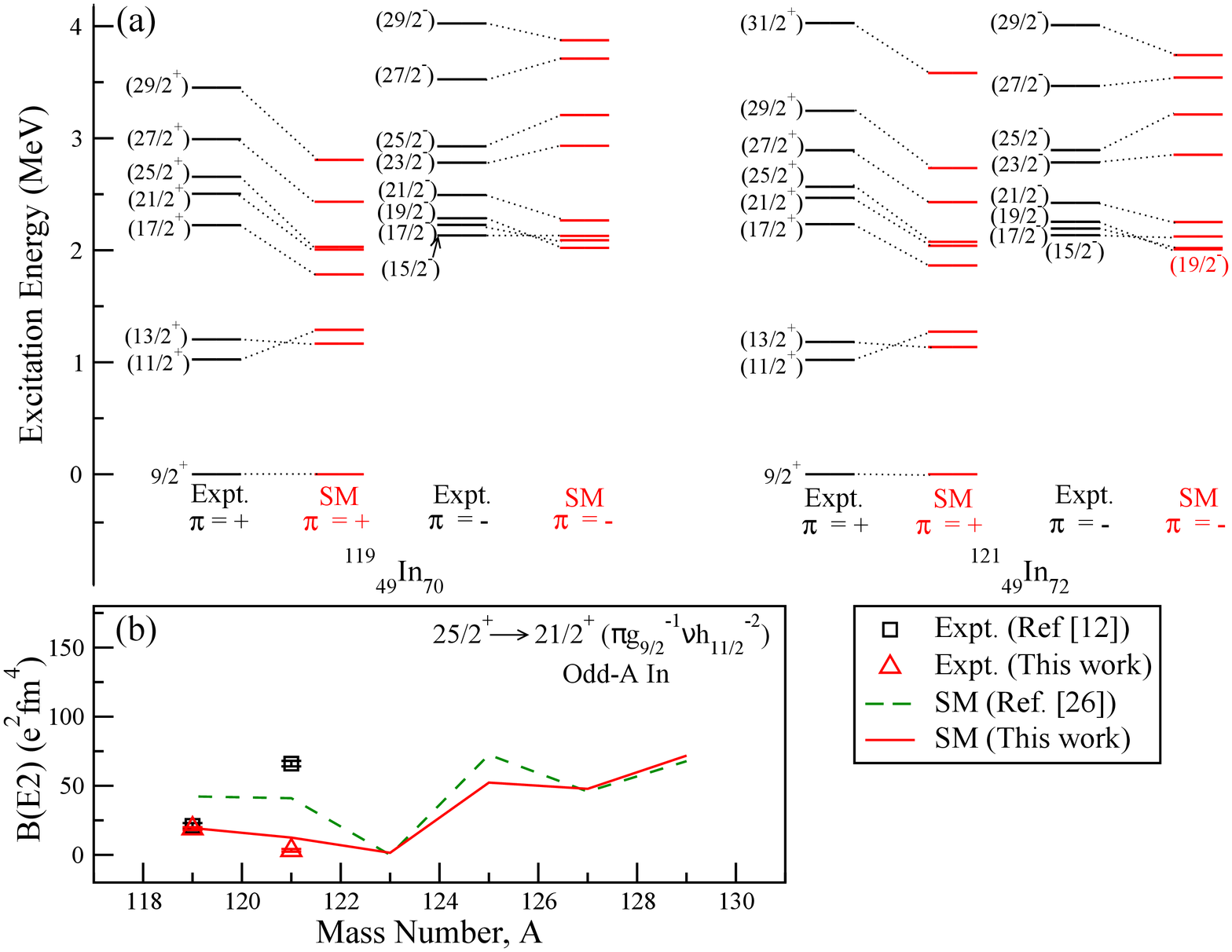}
\caption{\label{fig:dis_fig1} (Color online) (a) Comparison of the experimental (Expt.) level schemes with those obtained from shell model calculations using the interaction (SM) in the present work for both positive- and negative-parity states in $^{119,121}$In. The different states with the same spin are joined by dotted lines. (b) The experimental $B(E2)$ values for $25/2^{+} \rightarrow 21/2^{+}$ in odd-$A$ $^{119,121}$In from the previous work (Ref.~\cite{lu02}) and the present work are shown by the black square and red triangle respectively. The shell model (SM) calculations from Ref.~\cite{sb19} and the present work are shown by green dashed and red solid lines, respectively for $^{119-129}$In isotopes.}
\end{figure*}

Figure~\ref{fig:121in_fig3}(a) shows the half-life fit (one-component) for the decay spectrum upon gating on $99$~keV transition yielding a value of $T_{1/2} = 7.3(2)~\mu$s for the ($25/2^{+}$) state (in disagreement with the value of $350(50)$~ns reported in Ref.~\cite{lu02} and $17(2)~\mu$s reported in Ref.~\cite{re10}), as mentioned in the introduction. The newly reported half-life yielded $B(E2; 25/2^{+} \rightarrow 21/2^{+}) = 3.3(1)$~e$^{2}$fm$^{4}$. Similar fits were carried out for the other delayed transitions in the same cascade, namely the $214$, $953$, and $1181$~keV transitions yielding $7.5(2)~\mu$s, $7.0(2)~\mu$s, and $7.2(2)~\mu$s as shown in Figs.~\ref{fig:121in_fig3}(b), (c), and (d), respectively. That the half-life of the ($25/2^{+}$) state cannot be $350(50)$~ns (Ref.~\cite{lu02}), is also mentioned in Ref.~\cite{re10} and evident from our measurement. However, the disagreement between the present work and Ref.~\cite{re10} cannot be firmly understood as the present data set confirms the measurement of long-lived isomers in isotopes of Sb (Ref.~\cite{sb19}) and thus a long half-life of $17(2)~\mu$s could not be seen in the case of $^{121}$In. 

Further, it was not possible to obtain a $\gamma_{D}$ spectra with a very short time gate containing only the $953$ and $1181$~keV transitions. This suggests that the ($19/2^{-}$) state has a short half-life, which is below the sensitivity of the present setup, similar to what is observed in $^{119}$In. An estimate for the half-life could not be carried out in this case as the half-life of the ($25/2^{+}$) state is relatively large.

\section{\label{sec:Dis}Discussion}

Large-scale shell-model calculations, with a model space consisting of valence neutrons ($\nu$) in $d_{3/2}$, $s_{1/2}$ and $h_{11/2}$ and protons ($\pi$) in $p_{1/2}$, $g_{9/2}$ and $g_{7/2}$ orbitals, near the Fermi surface, were carried out to understand the structure of the high-spin states in neutron-rich $^{119,121}$In isotopes. The isotones of In and Sb ranging from $N=70$ to $N=80$ were also studied.  An interaction, derived from the {\it jj45pn} and {\it jj55pn} interactions~\cite{hj95} was adjusted to account for the missing correlations in the restricted model space used in this work to improve the agreement with both the level energies and $B(E2)$ values in Sn and Sb isotopes~\cite{sb19}. This interaction was then used, in the present work, to explain the energies and the $B(E2)$ values in the In isotopes. The calculations were performed using the NATHAN code~\cite{ca99}.  The theoretical $B(E2; 25/2^{+} \rightarrow 21/2^{+})~=~42.3$ and $41.0$~e$^{2}$fm$^{4}$ values for $^{119}$In and $^{121}$In, respectively, obtained from this interaction are not in agreement with the corresponding experimental values (the green dashed line in Fig.~\ref{fig:dis_fig1}(b)). The reason behind this could be attributed to the fact that in the Ref.~\cite{sb19}, no changes were made in the $\langle \pi g_{9/2} \nu h_{11/2};I \arrowvert \hat{\mathcal{H}}\arrowvert \pi g_{9/2} \nu h_{11/2};I\rangle$ two-body matrix elements of residual interaction. To improve the calculated values of the B(E2)'s in the In isotopes, the $\langle \pi g_{9/2} \nu h_{11/2};I \arrowvert \hat{\mathcal{H}}\arrowvert \pi g_{9/2} \nu h_{11/2};I\rangle$ ($I = 6, 8, 10$) matrix were modified by -200, -400 and +500 keV, respectively.\\
\begin{figure*}[]
\includegraphics[width=2.1\columnwidth]{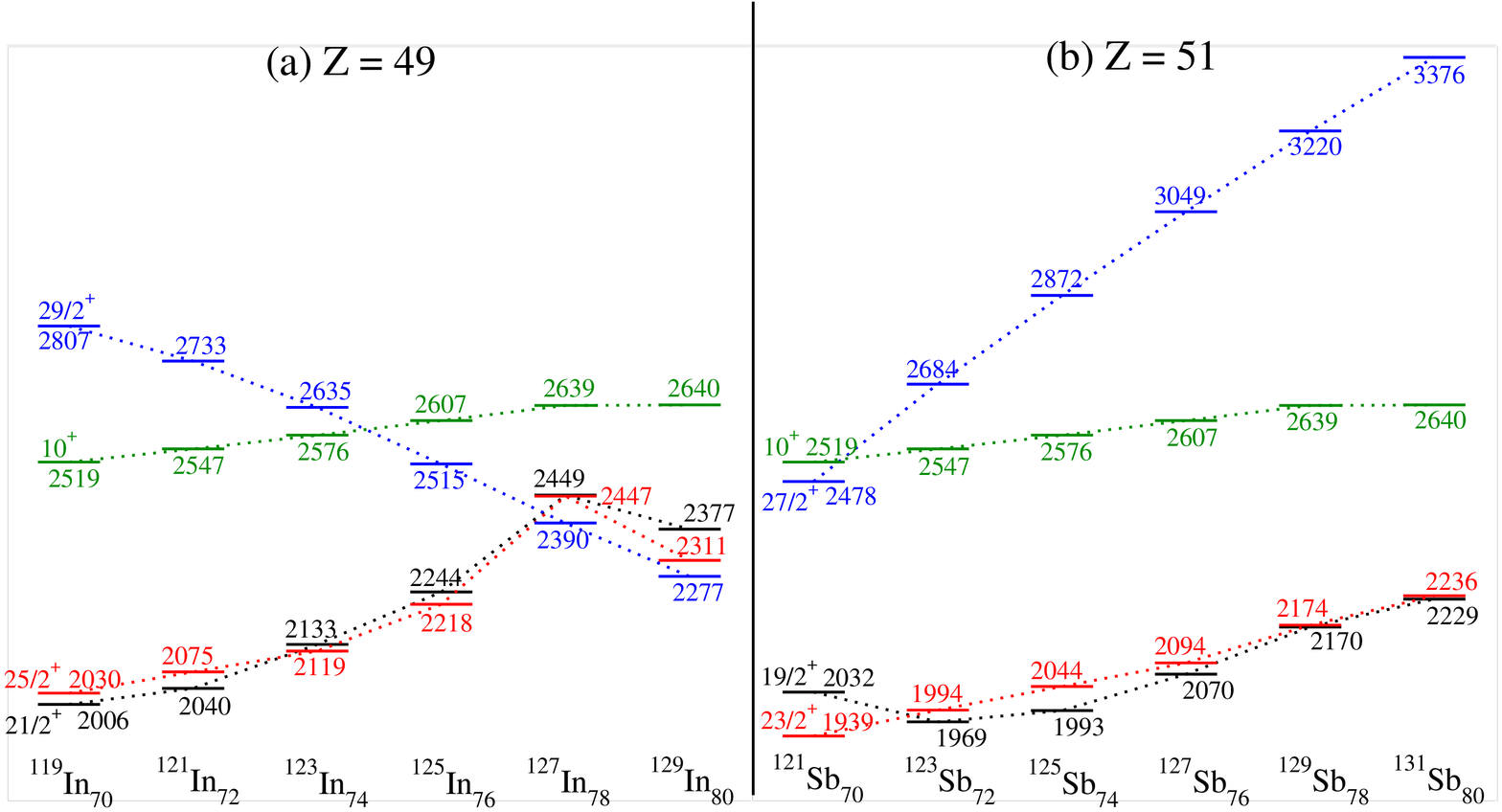} 
\includegraphics[width=2.1\columnwidth]{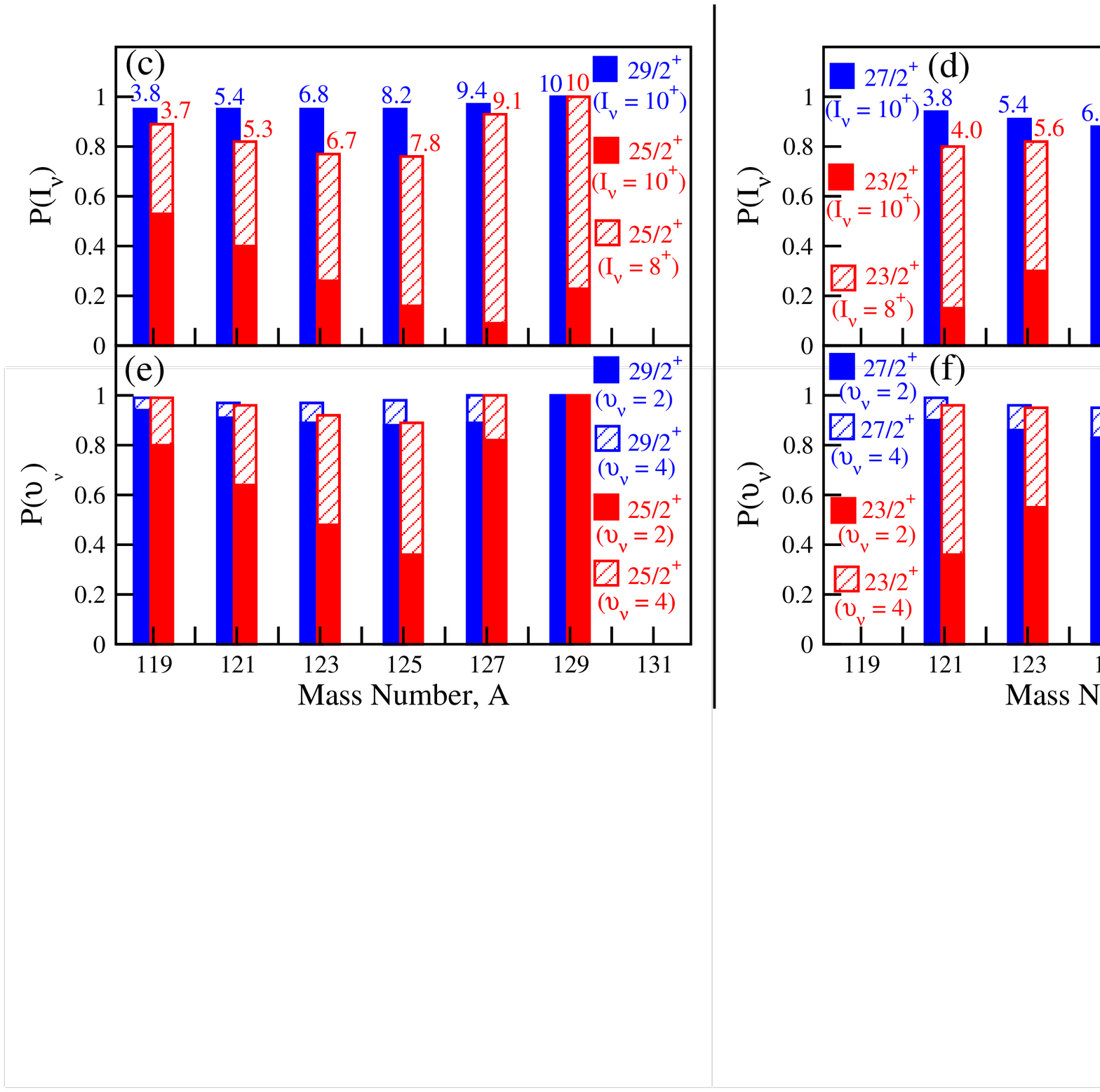}
\caption{\label{fig:dis_fig2} (Color online) Evolution of the calculated (SM) energies and the probabilities for the neutron angular momentum and neutron seniority for the, spin-orbit partners of In and Sb isotopes (a) Energies of $21/2^{+}$ (black), $25/2^{+}$ (red), and $29/2^{+}$ (blue) states in $Z = 49$ $^{119-129}$In isotopes; (b) Energies of the $19/2^{+}$ (black), $23/2^{+}$ (red), and $27/2^{+}$ (blue) level in $Z = 51$ $^{121-131}$Sb isotopes. The evolution of the $10^{+}$ states in $Z = 50$ $^{120-130}$Sn isotopes are also shown in green in these panels. The probability of the neutron angular momentum, $P(I_{\nu} = 8^{+}$ and $10^{+})$ for the (c) $29/2^{+}$ (blue filled), $25/2^{+}$ levels (red hatched filled) in In; and (d) for the $27/2^{+}$ (blue filled), $23/2^{+}$ (red filled and hatched) states in Sb isotopes. The occupancy of the $\nu h_{11/2}$ orbital for each level is also shown on top of each bar in these panels. The probability of the neutron seniority, $P(\upsilon_{\nu} = 2$ and $4)$ for the (e) $29/2^{+}$ (blue), $25/2^{+}$ states (red) in In; and (f) $27/2^{+}$ (blue filled and hatched), $23/2^{+}$ (red filled hatched) states in Sb isotopes.}
\end{figure*} 

Figure~\ref{fig:dis_fig1} shows the experimental excitation energies for the positive-parity and negative-parity levels (panel (a)) and $B(E2; 25/2^{+} \rightarrow 21/2^{+})$ values (panel (b)) in $^{119,121}$In, along with the calculations (SM) using the above mentioned modified interaction. It can be seen in Fig.~\ref{fig:dis_fig1}(a) that the experimental energies are reproduced within $\sim 500$~keV. The newly calculated $B(E2; 25/2^{+} \rightarrow 21/2^{+}) = 19.5$ and $12.5$~e$^{2}$fm$^{4}$ values (red solid line) were obtained for $^{119}$In and $^{121}$In isotopes respectively, are in reasonably good agreement with the present experimental values of $19(1)$ and $3.3(1)$~e$^{2}$fm$^{4}$ (red triangle), respectively, as shown in Fig.~\ref{fig:dis_fig1}(b). The shell model calculations show that the $25/2^{+}$ levels in odd-$A$ In have a dominant $\pi g_{9/2}^{-1} \nu h_{11/2}^{-2}$ configuration. The $B(E2; 25/2^{+} \rightarrow 21/2^{+})$ in the odd-$A$ In isotopes are expected to follow a similar trend as for the $B(E2; 10^{+} \rightarrow 8^{+})$ in even-$A$ Sn ($\nu h_{11/2}^{-2}$) and $B(E2; 23/2^{+} \rightarrow 19/2^{+})$ in odd-$A$ Sb ($\pi g_{7/2} \nu h_{11/2}^{-2}$) isotopes. Both these levels involve the ($\nu h_{11/2}^{-2}$) configuration. It was shown in Fig.~17 of Ref.~\cite{sb19} that both the experimental and the calculated $B(E2; 10^{+} \rightarrow 8^{+})$ in even-$A$ Sn and $B(E2; 23/2^{+} \rightarrow 19/2^{+})$ in odd-$A$ Sb show a parabolic behavior with a minimum around $N = 72, 74$. The experimental $B(E2;25/2^{+} \rightarrow 21/2^{+})$ values for the $^{119,121}$In isotopes reported in Ref.~\cite{lu02} show an increase (black squares in Fig.~\ref{fig:dis_fig1}(b)). However, the present experimental values (red triangles in Fig.~\ref{fig:dis_fig1}(b)) follow the expected behavior inherited from the Sn isotopes.

Based on a good agreement of the shell model results and the present experimental data, a systematic shell model analysis, using the interaction discussed above, was made for the isotopic chains of $^{121-131}$Sb and $^{119-129}$In.  The results of the calculations are shown in Fig.~\ref{fig:dis_fig2}.  The $10^{+}$ seniority isomer in even-$A$ $^{120-130}$Sn isotopes is known to be dominantly arising from $\nu h_{11/2}^{-2}$ configuration~\cite{fo81, pi11, as12, is14}.  Stretched angular momentum coupling of $\pi g_{7/2}$ particle in $^{121-131}$Sb ($\pi g_{9/2}$ hole in $^{119-129}$In) to the $\nu h_{11/2}^{-2}$ configuration leads to a maximum aligned spin of $I^{\pi}_{Max(\rm{Sb})} = 27/2^{+}$ in Sb and $I^{\pi}_{Max(\rm{In})} = 29/2^{+}$ in In. In the case of Sb, $I^{\pi}_{Max(\rm{Sb})} -2 = 23/2^{+}$ isomers were reported for $^{123-131}$Sb isotopes~\cite{sb19}.  However, in the case of In, $I^{\pi}_{Max(\rm{In})} -2 = 25/2^{+}$ isomers are observed only in the lighter odd-$A$ $^{119,121,125}$In isotopes, while $I^{\pi}_{Max(\rm{In})} =29/2^{+}$ isomers are reported in heavier odd-$A$ $^{127,129}$In isotopes~\cite{lu02, sc04}. The evolution of the energy of levels involved can be used to explain these observations. The results of shell model calculations for the $29/2^+$, $25/2^{+}$ and $21/2^{+}$ levels in $^{119-129}$In are shown in Fig.~\ref{fig:dis_fig2}(a). In Fig.~\ref{fig:dis_fig2}(b) the $27/2^+$, $23/2^{+}$ and $19/2^{+}$ levels in $^{121-131}$Sb are shown. Panels (a) and (b) include also the calculated $10^+$ state in $^{120-130}$Sn.  The energy of $10^+$ levels in Sn, associated with the $\nu h_{11/2}^{-2}$ configuration, is calculated to be relatively constant.  In In, the inversion of the $25/2^+$ and $21/2^{+}$ levels beyond $^{121}$In, is consistent with the reported long-lived $25/2^{+}$ isomer that does not decay to ${21/2}^+$ level in $^{123-129}$In~\cite{sc04}. Furthermore, the energy of the high lying ${29/2}^+$ level decreases steeply as the neutron number increases, it lies below the ${21/2}^+$ and ${25/2}^+$ in $^{127,129}$In. A similar conclusion was obtained from shell model calculations in Ref.~\cite{sc04} for $^{127,129}$In isotopes showing that the energy of the $29/2^{+}$ is lower than the $25/2^{+}$ level due to strong $\nu\pi$ interaction in the $(\pi g_{9/2}^{-1} \nu h_{11/2}^{-1}; {10^{-}})$ state. In the case of the Sb (Fig.~\ref{fig:dis_fig2}(b)), the calculations for the $^{121-131}$Sb isotopes show that the ordering of the $27/2^{+}$, $23/2^{+}$ and $19/2^{+}$ level is preserved, in agreement with the experimental observations.  An opposite behavior of the $I^{\pi}_{Max(\rm{In})} = 29/2^{+}$ in In and $I^{\pi}_{Max(\rm{Sb})} = 27/2^{+}$ in Sb can be clearly seen from the Fig.~\ref{fig:dis_fig2}(a) and (b): while the energy of the $29/2^+$ levels in In is decreasing with increasing neutron number, the energy of the $27/2^{+}$ levels in Sb increases.  This could be understood as a consequence of the hole-hole and particle-hole nature of the $\nu\pi$ interaction. At high particle occupancy of the $\nu h_{11/2}$ orbital the $\pi g \nu h_{11/2}$ interaction is of the hole-hole nature in In and of the particle-hole nature in Sb while at low particle occupancy of the $\nu h_{11/2}$ orbital it is of the hole-particle nature in In and particle-particle in Sb.

Beyond the energy levels, an in-depth analysis of the probability of the neutron angular momentum decomposition, $P(I_{\nu} = 8^{+}$ and $10^{+})$ and neutron seniority $P(\upsilon_{\nu} = 2$ and $4)$, for the $29/2^{+}$ and $25/2^{+}$ states in $^{119-129}$In; $27/2^{+}$ and $23/2^{+}$ states in $^{121-131}$Sb, using the shell model calculations, was made. These probability distributions are shown in Figs.~\ref{fig:dis_fig2}(c,d,e,f), respectively.  Figure~\ref{fig:dis_fig2}(c) shows that for $^{119-129}$In, the $29/2^{+}$ level has a dominant contribution from $I_{\nu} = 10^{+}$ ($I_{\nu} = 8^{+}$ is not possible for this spin).  The contribution from $I_{\nu} = 10^{+}$ to the $29/2^{+}$ level reaches the unity only in $^{129}$In, in the $^{119-127}$In isotopes there is an admixture arising from $I_{\nu} > 10^{+}$ with $\upsilon_{\nu} = 4$ as can be seen in Fig~\ref{fig:dis_fig2}(e).  For the $25/2^{+}$ state in $^{119-129}$In, both the contributions from $I_{\nu} = 8^{+}$ and $10^{+}$ are shown by red hatched and red filled bars. It can be seen from Fig~\ref{fig:dis_fig2}(c) that the contribution from $I_{\nu} = 8^{+}$ dominates at high mass number and that the contribution from $I_{\nu} = 10^{+}$ is increasing with decreasing mass number.  At lower mass number the contribution to the $25/2^{+}$ state form the $\upsilon_{\nu} = 2$ dominates, while with increasing mass number $\upsilon_{\nu} = 4$ increases. This trend suddenly breaks beyond $A=125$, since at $A=129$ $\upsilon_{\nu} = 4$ is no longer possible. This could be also a reason of the presence of the discontinuity of the increasing energy trend of the $21/2^+$ and $25/2^+$ with increasing mass number (see Fig.~\ref{fig:dis_fig2}(c)).

A similar decomposition is shown in Figs.~\ref{fig:dis_fig2}(d) and (f) for Sb isotopes. Figure~\ref{fig:dis_fig2} (d) shows that for $^{121-131}$Sb, the $27/2^{+}$ state also has a dominant contribution from $I_{\nu} = 10^{+}$.  For the $23/2^{+}$ levels in $^{121-131}$Sb, both the contributions from $I_{\nu} = 8^{+}$ and $10^{+}$ are shown by red hatched and red filled bars respectively. It can be seen from Fig.~\ref{fig:dis_fig2} (d) that on the contrary to In, the contribution from $I_{\nu} = 10^{+}$ dominates at high mass number and that the contribution of $I_{\nu} = 8^{+}$ increases with decreasing mass number. Similarly in Fig~\ref{fig:dis_fig2}(f) the inverted trends, as compared to $25/2^+ $states in In, can be seen in the contribution of $\upsilon_{\nu} = 2$ and $4$ to the $23/2^+$ states.

Summarizing, the shell model calculations, the presence of the $25/2^{+}$ isomers in lighter In isotopes was reproduced and the $B(E2)$ values of the corresponding $\gamma$-decay were well described.  An in-depth analysis and the comparison of the $I^{\pi}_{Max(\rm{In})} = 29/2^{+}$ and $I^{\pi}_{Max(\rm{In})} -2 = 25/2^{+}$ in In and $I^{\pi}_{Max(\rm{Sb})} = 27/2^{+}$ and $I^{\pi}_{Max(\rm{Sb})} -2 = 23/2^{+}$ in Sb revealed several ``mirror''-like symmetries between the corresponding states in In and Sb, as a function of the mass number:
\begin{itemize}
\item The energy of the $I^{\pi}_{Max(\rm{In})} = 29/2^{+}$ is strongly down sloping while the energy of the $I^{\pi}_{Max(\rm{Sb})} = 27/2^{+}$ is up sloping, with increasing mass number. Both states result dominantly from the stretched coupling of proton-hole or proton-particle with the neutron $10^+$ with seniority $\upsilon_{\nu} = 2$ state of the underlying Sn core.
\item The energies of the $I^{\pi}_{Max(\rm{In})} -2 = 25/2^{+}$ and the $I^{\pi}_{Max(\rm{Sb})} -2 = 23/2^{+}$ are similar and increase slightly with increasing mass number. Both result dominantly from the coupling of proton-hole or proton-particle coupled to a mixture of the neutron $8^+$ and $10^+$ states with strong mixing of seniorities $\upsilon_{\nu} = 2$ and $4$. In In, at lower mass number the dominant contribution is from the $10^+$ state and the seniority $\upsilon_{\nu} = 2$, both decrease with increasing mass number. On the contrary, in Sb, at lower mass number the dominant contribution is from the $8^+$ state and the seniority $\upsilon_{\nu} = 4$, and the trends are opposite to those observed in In.
\end{itemize}
Those ``mirror''-like symmetries could be understood as resulting from the particle-hole and the hole-hole symmetry of the $\nu\pi$ interaction.

\section{\label{sec:SumCon}Summary and Conclusions}

The neutron-rich odd-$A$ $^{119,121}$In isotopes were produced as fission fragments in the reaction $^{9}$Be ($^{238}$U, f) at energies around the Coulomb barrier. New prompt transitions were identified above the $25/2^{+}$ isomer in $^{121}$In along with the remeasurement of the half-life of this isomer. Additionally, the possibility of a short half-life of the ($19/2^{-}$) isomeric states in odd-$A$ $^{119,121}$In isotopes was demonstrated. These results were possible using the unique combination of AGATA, VAMOS++ and EXOGAM detectors, measuring the prompt-delayed spectroscopy of isotopically identified fission fragments. Shell model calculations, using a modification of the interaction used in Ref.~\cite{sb19}, were used to interpret the data and also study the evolution of the energy of the $21/2^{-}$, $25/2^{-}$ and $29/2^{-}$ levels in isotopic chains of odd-$A$ In and the corresponding $19/2^{-}$; $23/2^{-}$ and $25/2^{-}$ levels in odd-$A$ Sb, using a decomposition in terms of the probability distributions of the neutron angular momentum and neutron seniority.  A good agreement was obtained between the measured energies and the $B(E2)$ transition probabilities and shell model calculations. In addition, in the odd-$A$ In isotopes (having a $\pi g_{9/2}^{-1} \nu h_{11/2}^{-2}$ configuration) the newly measured $B(E2; 25/2^{+} \rightarrow 21/2^{+})$ were shown to follow a similar behavior to those for the $B(E2; 10^{+} \rightarrow 8^{+})$ in even-$A$ Sn ($\nu h_{11/2}^{-2}$) isotopes and $B(E2; 23/2^{+} \rightarrow 19/2^{+})$ in odd-$A$ Sb isotopes ($\pi g_{7/2}^{1} \nu h_{11/2}^{-2}$). Shell model calculations presented in this work show that the $\pi g_{9/2}$ hole coupled to the $\nu h_{11/2}^{-2}$ configuration leads to the lowering of the energy of the maximum aligned spin $29/2^{+}$ level as a function of increasing neutron number, thus leading to the observation of $29/2^{+}$ isomeric states in the $^{127,129}$In isotopes.  The $\pi g_{7/2}$ particle coupling to the $\nu h_{11/2}^{-2}$ configuration leads to the increase of the energy of maximum aligned spin $27/2^{+}$ level as a function of increasing neutron number leading to $23/2^{+}$ isomeric states in $^{123-131}$Sb isotopes. These results illustrate the role of the particle-hole and the hole-hole symmetry of the $\nu\pi$ interaction. Future experiments using  prompt-delayed spectroscopy that are sensitive to short lifetimes could allow to determine the lifetime of the $19/2^-$ state in $^{119,121}$In  that would further clarify  the high spin structure of these isotopes.

\section{\label{sec:Ack}Acknowledgments}

The authors would like to thank the AGATA Collaboration for the availability of the AGATA $\gamma$-ray tracking array at GANIL. We acknowledge the important technical contributions of GANIL accelerator staff.  We thank C. Schmitt for help during the experiment and careful reading of the manuscript. We acknowledge A. O. Macchiavelli for help during the experiment. We also thank P. Van Isacker for his valuable discussions on the theoretical interpretation and careful reading of the manuscript. PB and AM acknowledge support from the Polish National Science Centre (NCN) under Contract No. 2016/22/M/ST2/00269 and the French LEA COPIGAL project. SBi, RB, SBh, SBh and RP acknowledge support from CEFIPRA project No. 5604-4 and the LIA France-India agreement. HLC and PF acknowledge support from the U.S. Department of Energy, Office of Science, Office of Nuclear Physics under Contract No. DE-AC02-05CH11231 (LBNL). RMPV acknowledge partial support by Ministry of Science, Spain, under the grants BES-2012-061407, SEV-2014-0398, FPA2017-84756-C4 and by EU FEDER funds.

 \bibliography{BiblioIn.bib} 

\begin{thebibliography}{29}%
\makeatletter
\providecommand \@ifxundefined [1]{%
 \@ifx{#1\undefined}
}%
\providecommand \@ifnum [1]{%
 \ifnum #1\expandafter \@firstoftwo
 \else \expandafter \@secondoftwo
 \fi
}%
\providecommand \@ifx [1]{%
 \ifx #1\expandafter \@firstoftwo
 \else \expandafter \@secondoftwo
 \fi
}%
\providecommand \natexlab [1]{#1}%
\providecommand \enquote  [1]{``#1''}%
\providecommand \bibnamefont  [1]{#1}%
\providecommand \bibfnamefont [1]{#1}%
\providecommand \citenamefont [1]{#1}%
\providecommand \href@noop [0]{\@secondoftwo}%
\providecommand \href [0]{\begingroup \@sanitize@url \@href}%
\providecommand \@href[1]{\@@startlink{#1}\@@href}%
\providecommand \@@href[1]{\endgroup#1\@@endlink}%
\providecommand \@sanitize@url [0]{\catcode `\\12\catcode `\$12\catcode
  `\&12\catcode `\#12\catcode `\^12\catcode `\_12\catcode `\%12\relax}%
\providecommand \@@startlink[1]{}%
\providecommand \@@endlink[0]{}%
\providecommand \url  [0]{\begingroup\@sanitize@url \@url }%
\providecommand \@url [1]{\endgroup\@href {#1}{\urlprefix }}%
\providecommand \urlprefix  [0]{URL }%
\providecommand \Eprint [0]{\href }%
\providecommand \doibase [0]{http://dx.doi.org/}%
\providecommand \selectlanguage [0]{\@gobble}%
\providecommand \bibinfo  [0]{\@secondoftwo}%
\providecommand \bibfield  [0]{\@secondoftwo}%
\providecommand \translation [1]{[#1]}%
\providecommand \BibitemOpen [0]{}%
\providecommand \bibitemStop [0]{}%
\providecommand \bibitemNoStop [0]{.\EOS\space}%
\providecommand \EOS [0]{\spacefactor3000\relax}%
\providecommand \BibitemShut  [1]{\csname bibitem#1\endcsname}%
\let\auto@bib@innerbib\@empty
\bibitem [{\citenamefont {Mayer}(1949)}]{ma49}%
  \BibitemOpen
  \bibfield  {author} {\bibinfo {author} {\bibfnamefont {M.~G.}\ \bibnamefont
  {Mayer}},\ }\href {\doibase 10.1103/PhysRev.75.1969} {\bibfield  {journal}
  {\bibinfo  {journal} {Phys. Rev.}\ }\textbf {\bibinfo {volume} {75}},\
  \bibinfo {pages} {1969} (\bibinfo {year} {1949})}\BibitemShut {NoStop}%
\bibitem [{\citenamefont {Haxel}\ \emph {et~al.}(1949)\citenamefont {Haxel},
  \citenamefont {Jensen},\ and\ \citenamefont {Suess}}]{ha49}%
  \BibitemOpen
  \bibfield  {author} {\bibinfo {author} {\bibfnamefont {O.}~\bibnamefont
  {Haxel}}, \bibinfo {author} {\bibfnamefont {J.~H.~D.}\ \bibnamefont
  {Jensen}}\ and\ \bibinfo {author} {\bibfnamefont {H.~E.}\ \bibnamefont
  {Suess}},\ }\href {\doibase 10.1103/PhysRev.75.1766.2} {\bibfield  {journal}
  {\bibinfo  {journal} {Phys. Rev.}\ }\textbf {\bibinfo {volume} {75}},\
  \bibinfo {pages} {1766} (\bibinfo {year} {1949})}\BibitemShut {NoStop}%
\bibitem [{\citenamefont {Fogelberg}\ \emph {et~al.}(1981)\citenamefont
  {Fogelberg}, \citenamefont {Heyde},\ and\ \citenamefont {Sau}}]{fo81}%
  \BibitemOpen
  \bibfield  {author} {\bibinfo {author} {\bibfnamefont {B.}~\bibnamefont
  {Fogelberg}}, \bibinfo {author} {\bibfnamefont {K.}~\bibnamefont {Heyde}}\
  and\ \bibinfo {author} {\bibfnamefont {J.}~\bibnamefont {Sau}},\ }\href
  {\doibase 10.1016/0375-9474(81)90374-2} {\bibfield  {journal} {\bibinfo
  {journal} {Nucl. Phys. A}\ }\textbf {\bibinfo {volume} {352}},\ \bibinfo
  {pages} {157} (\bibinfo {year} {1981})}\BibitemShut {NoStop}%
\bibitem [{\citenamefont {Pietri}\ \emph {et~al.}(2011)\citenamefont {Pietri},
  \citenamefont {Jungclaus}, \citenamefont {G\'orska}, \citenamefont {Grawe},
  \citenamefont {Pf\"utzner}, \citenamefont {C\'aceres}, \citenamefont
  {Detistov}, \citenamefont {Lalkovski}, \citenamefont {Modamio}, \citenamefont
  {Podoly\'ak}, \citenamefont {Regan}, \citenamefont {Rudolph}, \citenamefont
  {Walker}, \citenamefont {Werner-Malento}, \citenamefont {Bednarczyk},
  \citenamefont {Doornenbal}, \citenamefont {Geissel}, \citenamefont {Gerl},
  \citenamefont {Grebosz}, \citenamefont {Kojouharov}, \citenamefont {Kurz},
  \citenamefont {Prokopowicz}, \citenamefont {Schaffner}, \citenamefont
  {Wollersheim}, \citenamefont {Andgren}, \citenamefont {Benlliure},
  \citenamefont {Benzoni}, \citenamefont {Bruce}, \citenamefont {Casarejos},
  \citenamefont {Cederwall}, \citenamefont {Crespi}, \citenamefont {Hadinia},
  \citenamefont {Hellstr\"om}, \citenamefont {Hoischen}, \citenamefont {Ilie},
  \citenamefont {Khaplanov}, \citenamefont {Kmiecik}, \citenamefont {Kumar},
  \citenamefont {Maj}, \citenamefont {Mandal}, \citenamefont {Montes},
  \citenamefont {Myalski}, \citenamefont {Simpson}, \citenamefont {Steer},
  \citenamefont {Tashenov},\ and\ \citenamefont {Wieland}}]{pi11}%
  \BibitemOpen
  \bibfield  {author} {\bibinfo {author} {\bibfnamefont {S.}~\bibnamefont
  {Pietri}}, \bibinfo {author} {\bibfnamefont {A.}~\bibnamefont {Jungclaus}},
  \bibinfo {author} {\bibfnamefont {M.}~\bibnamefont {G\'orska}}, \bibinfo
  {author} {\bibfnamefont {H.}~\bibnamefont {Grawe}}, \bibinfo {author}
  {\bibfnamefont {M.}~\bibnamefont {Pf\"utzner}}, \bibinfo {author}
  {\bibfnamefont {L.}~\bibnamefont {C\'aceres}}, \bibinfo {author}
  {\bibfnamefont {P.}~\bibnamefont {Detistov}}, \bibinfo {author}
  {\bibfnamefont {S.}~\bibnamefont {Lalkovski}}, \bibinfo {author}
  {\bibfnamefont {V.}~\bibnamefont {Modamio}}, \bibinfo {author} {\bibfnamefont
  {Z.}~\bibnamefont {Podoly\'ak}}, \bibinfo {author} {\bibfnamefont {P.~H.}\
  \bibnamefont {Regan}}, \bibinfo {author} {\bibfnamefont {D.}~\bibnamefont
  {Rudolph}},  \emph {et~al.},\ }\href {\doibase 10.1103/PhysRevC.83.044328}
  {\bibfield  {journal} {\bibinfo  {journal} {Phys. Rev. C}\ }\textbf {\bibinfo
  {volume} {83}},\ \bibinfo {pages} {044328} (\bibinfo {year}
  {2011})}\BibitemShut {NoStop}%
\bibitem [{\citenamefont {Astier}\ \emph {et~al.}(2012)\citenamefont {Astier},
  \citenamefont {Porquet}, \citenamefont {Theisen}, \citenamefont {Verney},
  \citenamefont {Deloncle}, \citenamefont {Houry}, \citenamefont {Lucas},
  \citenamefont {Azaiez}, \citenamefont {Barreau}, \citenamefont {Curien},
  \citenamefont {Dorvaux}, \citenamefont {Duch\^ene}, \citenamefont {Gall},
  \citenamefont {Redon}, \citenamefont {Rousseau},\ and\ \citenamefont
  {St\'ezowski}}]{as12}%
  \BibitemOpen
  \bibfield  {author} {\bibinfo {author} {\bibfnamefont {A.}~\bibnamefont
  {Astier}}, \bibinfo {author} {\bibfnamefont {M.-G.}\ \bibnamefont {Porquet}},
  \bibinfo {author} {\bibfnamefont {C.}~\bibnamefont {Theisen}}, \bibinfo
  {author} {\bibfnamefont {D.}~\bibnamefont {Verney}}, \bibinfo {author}
  {\bibfnamefont {I.}~\bibnamefont {Deloncle}}, \bibinfo {author}
  {\bibfnamefont {M.}~\bibnamefont {Houry}}, \bibinfo {author} {\bibfnamefont
  {R.}~\bibnamefont {Lucas}}, \bibinfo {author} {\bibfnamefont
  {F.}~\bibnamefont {Azaiez}}, \bibinfo {author} {\bibfnamefont
  {G.}~\bibnamefont {Barreau}}, \bibinfo {author} {\bibfnamefont
  {D.}~\bibnamefont {Curien}}, \bibinfo {author} {\bibfnamefont
  {O.}~\bibnamefont {Dorvaux}}, \bibinfo {author} {\bibfnamefont
  {G.}~\bibnamefont {Duch\^ene}},  \emph {et~al.},\ }\href {\doibase
  10.1103/PhysRevC.85.054316} {\bibfield  {journal} {\bibinfo  {journal} {Phys.
  Rev. C}\ }\textbf {\bibinfo {volume} {85}},\ \bibinfo {pages} {054316}
  (\bibinfo {year} {2012})}\BibitemShut {NoStop}%
\bibitem [{\citenamefont {Iskra}\ \emph {et~al.}(2014)\citenamefont {Iskra},
  \citenamefont {Broda}, \citenamefont {Wrzesinski}, \citenamefont {Carpenter},
  \citenamefont {Chiara}, \citenamefont {Fornal}, \citenamefont {Hoteling},
  \citenamefont {Janssens}, \citenamefont {Kondev}, \citenamefont
  {Kr{\'{o}}las}, \citenamefont {Lauritsen}, \citenamefont {Pawat},
  \citenamefont {Seweryniak}, \citenamefont {Stefanescu}, \citenamefont
  {Walters},\ and\ \citenamefont {Zhu}}]{is14}%
  \BibitemOpen
  \bibfield  {author} {\bibinfo {author} {\bibfnamefont {W.}~\bibnamefont
  {Iskra}}, \bibinfo {author} {\bibfnamefont {R.}~\bibnamefont {Broda}},
  \bibinfo {author} {\bibfnamefont {J.}~\bibnamefont {Wrzesinski}}, \bibinfo
  {author} {\bibfnamefont {M.~P.}\ \bibnamefont {Carpenter}}, \bibinfo {author}
  {\bibfnamefont {C.~J.}\ \bibnamefont {Chiara}}, \bibinfo {author}
  {\bibfnamefont {B.}~\bibnamefont {Fornal}}, \bibinfo {author} {\bibfnamefont
  {N.}~\bibnamefont {Hoteling}}, \bibinfo {author} {\bibfnamefont {R.~V.}\
  \bibnamefont {Janssens}}, \bibinfo {author} {\bibfnamefont {F.~G.}\
  \bibnamefont {Kondev}}, \bibinfo {author} {\bibfnamefont {W.}~\bibnamefont
  {Kr{\'{o}}las}}, \bibinfo {author} {\bibfnamefont {T.}~\bibnamefont
  {Lauritsen}}, \bibinfo {author} {\bibfnamefont {T.}~\bibnamefont {Pawat}},
  \emph {et~al.},\ }\href {\doibase 10.5506/APhysPolB.44.395} {\bibfield
  {journal} {\bibinfo  {journal} {Act. Phys. Pol. B}\ }\textbf {\bibinfo
  {volume} {89}},\ \bibinfo {pages} {395} (\bibinfo {year} {2014})}\BibitemShut
  {NoStop}%
\bibitem [{\citenamefont {Judson}\ \emph {et~al.}(2007)\citenamefont {Judson},
  \citenamefont {Bruce}, \citenamefont {Kib\'edi}, \citenamefont {Dracoulis},
  \citenamefont {Byrne}, \citenamefont {Lane}, \citenamefont {Maier},
  \citenamefont {Moon}, \citenamefont {Nieminen}, \citenamefont {Orce},\ and\
  \citenamefont {Taylor}}]{ju07}%
  \BibitemOpen
  \bibfield  {author} {\bibinfo {author} {\bibfnamefont {D.~S.}\ \bibnamefont
  {Judson}}, \bibinfo {author} {\bibfnamefont {A.~M.}\ \bibnamefont {Bruce}},
  \bibinfo {author} {\bibfnamefont {T.}~\bibnamefont {Kib\'edi}}, \bibinfo
  {author} {\bibfnamefont {G.~D.}\ \bibnamefont {Dracoulis}}, \bibinfo {author}
  {\bibfnamefont {A.~P.}\ \bibnamefont {Byrne}}, \bibinfo {author}
  {\bibfnamefont {G.~J.}\ \bibnamefont {Lane}}, \bibinfo {author}
  {\bibfnamefont {K.~H.}\ \bibnamefont {Maier}}, \bibinfo {author}
  {\bibfnamefont {C.-B.}\ \bibnamefont {Moon}}, \bibinfo {author}
  {\bibfnamefont {P.}~\bibnamefont {Nieminen}}, \bibinfo {author}
  {\bibfnamefont {J.~N.}\ \bibnamefont {Orce}}, \ and\ \bibinfo {author}
  {\bibfnamefont {M.~J.}\ \bibnamefont {Taylor}},\ }\href {\doibase
  10.1103/PhysRevC.76.054306} {\bibfield  {journal} {\bibinfo  {journal} {Phys.
  Rev. C}\ }\textbf {\bibinfo {volume} {76}},\ \bibinfo {pages} {054306}
  (\bibinfo {year} {2007})}\BibitemShut {NoStop}%
\bibitem [{\citenamefont {Watanabe}\ \emph
  {et~al.}(2009{\natexlab{a}})\citenamefont {Watanabe}, \citenamefont {Lane},
  \citenamefont {Dracoulis}, \citenamefont {Kib\'edi}, \citenamefont {Byrne},
  \citenamefont {Nieminen}, \citenamefont {Hughes}, \citenamefont {Kondev},
  \citenamefont {Carpenter}, \citenamefont {Janssens}, \citenamefont
  {Lauritsen}, \citenamefont {Seweryniak}, \citenamefont {Zhu}, \citenamefont
  {Chowdhury},\ and\ \citenamefont {Moon}}]{wa09}%
  \BibitemOpen
  \bibfield  {author} {\bibinfo {author} {\bibfnamefont {H.}~\bibnamefont
  {Watanabe}}, \bibinfo {author} {\bibfnamefont {G.~J.}\ \bibnamefont {Lane}},
  \bibinfo {author} {\bibfnamefont {G.~D.}\ \bibnamefont {Dracoulis}}, \bibinfo
  {author} {\bibfnamefont {T.}~\bibnamefont {Kib\'edi}}, \bibinfo {author}
  {\bibfnamefont {A.~P.}\ \bibnamefont {Byrne}}, \bibinfo {author}
  {\bibfnamefont {P.}~\bibnamefont {Nieminen}}, \bibinfo {author}
  {\bibfnamefont {R.~O.}\ \bibnamefont {Hughes}}, \bibinfo {author}
  {\bibfnamefont {F.~G.}\ \bibnamefont {Kondev}}, \bibinfo {author}
  {\bibfnamefont {M.~P.}\ \bibnamefont {Carpenter}}, \bibinfo {author}
  {\bibfnamefont {R.~V.~F.}\ \bibnamefont {Janssens}}, \bibinfo {author}
  {\bibfnamefont {T.}~\bibnamefont {Lauritsen}}, \bibinfo {author}
  {\bibfnamefont {D.}~\bibnamefont {Seweryniak}},  \emph {et~al.},\ }\href
  {\doibase 10.1103/PhysRevC.79.024306} {\bibfield  {journal} {\bibinfo
  {journal} {Phys. Rev. C}\ }\textbf {\bibinfo {volume} {79}},\ \bibinfo
  {pages} {024306} (\bibinfo {year} {2009}{\natexlab{a}})}\BibitemShut
  {NoStop}%
\bibitem [{\citenamefont {Watanabe}\ \emph
  {et~al.}(2009{\natexlab{b}})\citenamefont {Watanabe}, \citenamefont {Lane},
  \citenamefont {Dracoulis}, \citenamefont {Byrne}, \citenamefont {Nieminen},
  \citenamefont {Kondev}, \citenamefont {Ogawa}, \citenamefont {Carpenter},
  \citenamefont {Janssens}, \citenamefont {Lauritsen}, \citenamefont
  {Seweryniak}, \citenamefont {Zhu},\ and\ \citenamefont
  {Chowdhury}}]{wa09epja}%
  \BibitemOpen
  \bibfield  {author} {\bibinfo {author} {\bibfnamefont {H.}~\bibnamefont
  {Watanabe}}, \bibinfo {author} {\bibfnamefont {G.~J.}\ \bibnamefont {Lane}},
  \bibinfo {author} {\bibfnamefont {G.~D.}\ \bibnamefont {Dracoulis}}, \bibinfo
  {author} {\bibfnamefont {A.~P.}\ \bibnamefont {Byrne}}, \bibinfo {author}
  {\bibfnamefont {P.}~\bibnamefont {Nieminen}}, \bibinfo {author}
  {\bibfnamefont {F.~G.}\ \bibnamefont {Kondev}}, \bibinfo {author}
  {\bibfnamefont {K.}~\bibnamefont {Ogawa}}, \bibinfo {author} {\bibfnamefont
  {M.~P.}\ \bibnamefont {Carpenter}}, \bibinfo {author} {\bibfnamefont
  {R.~V.~F.}\ \bibnamefont {Janssens}}, \bibinfo {author} {\bibfnamefont
  {T.}~\bibnamefont {Lauritsen}}, \bibinfo {author} {\bibfnamefont
  {D.}~\bibnamefont {Seweryniak}}, \bibinfo {author} {\bibfnamefont
  {S.}~\bibnamefont {Zhu}},  \emph {et~al.},\ }\href {\doibase
  10.1140/epja/i2009-10881-7} {\bibfield  {journal} {\bibinfo  {journal} {Eur.
  Phys. J. A}\ }\textbf {\bibinfo {volume} {42}},\ \bibinfo {pages} {163}
  (\bibinfo {year} {2009}{\natexlab{b}})}\BibitemShut {NoStop}%
\bibitem [{\citenamefont {Genevey}\ \emph {et~al.}(2003)\citenamefont
  {Genevey}, \citenamefont {Pinston}, \citenamefont {Faust}, \citenamefont
  {Orlandi}, \citenamefont {Scherillo}, \citenamefont {Simpson}, \citenamefont
  {Tsekhanovich}, \citenamefont {Covello}, \citenamefont {Gargano},\ and\
  \citenamefont {Urban}}]{ge03}%
  \BibitemOpen
  \bibfield  {author} {\bibinfo {author} {\bibfnamefont {J.}~\bibnamefont
  {Genevey}}, \bibinfo {author} {\bibfnamefont {J.~A.}\ \bibnamefont
  {Pinston}}, \bibinfo {author} {\bibfnamefont {H.~R.}\ \bibnamefont {Faust}},
  \bibinfo {author} {\bibfnamefont {R.}~\bibnamefont {Orlandi}}, \bibinfo
  {author} {\bibfnamefont {A.}~\bibnamefont {Scherillo}}, \bibinfo {author}
  {\bibfnamefont {G.~S.}\ \bibnamefont {Simpson}}, \bibinfo {author}
  {\bibfnamefont {I.~S.}\ \bibnamefont {Tsekhanovich}}, \bibinfo {author}
  {\bibfnamefont {A.}~\bibnamefont {Covello}}, \bibinfo {author} {\bibfnamefont
  {A.}~\bibnamefont {Gargano}}, \ and\ \bibinfo {author} {\bibfnamefont
  {W.}~\bibnamefont {Urban}},\ }\href {\doibase 10.1103/PhysRevC.67.054312}
  {\bibfield  {journal} {\bibinfo  {journal} {Phys. Rev. C}\ }\textbf {\bibinfo
  {volume} {67}},\ \bibinfo {pages} {054312} (\bibinfo {year}
  {2003})}\BibitemShut {NoStop}%
\bibitem [{\citenamefont {{J. Genevey}}\ \emph {et~al.}(2000)\citenamefont {{J.
  Genevey}}, \citenamefont {{J.A. Pinston}}, \citenamefont {{H. Faust}},
  \citenamefont {{C. Foin}}, \citenamefont {{S. Oberstedt}},\ and\
  \citenamefont {{M. Rejmund}}}]{ge00}%
  \BibitemOpen
  \bibfield  {author} {\bibinfo {author} {\bibnamefont {{J. Genevey}}},
  \bibinfo {author} {\bibnamefont {{J.A. Pinston}}}, \bibinfo {author}
  {\bibnamefont {{H. Faust}}}, \bibinfo {author} {\bibnamefont {{C. Foin}}},
  \bibinfo {author} {\bibnamefont {{S. Oberstedt}}}, \ and\ \bibinfo {author}
  {\bibnamefont {{M. Rejmund}}},\ }\href {\doibase 10.1007/s100500070036}
  {\bibfield  {journal} {\bibinfo  {journal} {Eur. Phys. J. A}\ }\textbf
  {\bibinfo {volume} {9}},\ \bibinfo {pages} {191} (\bibinfo {year}
  {2000})}\BibitemShut {NoStop}%
\bibitem [{\citenamefont {Lucas}\ \emph {et~al.}(2002)\citenamefont {Lucas},
  \citenamefont {Porquet}, \citenamefont {Venkova}, \citenamefont {Deloncle},
  \citenamefont {Houry}, \citenamefont {Theisen}, \citenamefont {Astier},
  \citenamefont {Bauchet}, \citenamefont {Lalkovski}, \citenamefont {Barreau},
  \citenamefont {Buforn}, \citenamefont {Doan}, \citenamefont {Donadille},
  \citenamefont {Dorvaux}, \citenamefont {Durell}, \citenamefont {Ethvignot},
  \citenamefont {Gall}, \citenamefont {Grimwood}, \citenamefont {Korten},
  \citenamefont {Le~Coz}, \citenamefont {Meyer}, \citenamefont {Minkova},
  \citenamefont {Pr{\'e}vost}, \citenamefont {Redon}, \citenamefont {Roach},
  \citenamefont {Schulz}, \citenamefont {Smith}, \citenamefont
  {St{\'e}zowski},\ and\ \citenamefont {Varley}}]{lu02}%
  \BibitemOpen
  \bibfield  {author} {\bibinfo {author} {\bibfnamefont {R.}~\bibnamefont
  {Lucas}}, \bibinfo {author} {\bibfnamefont {M.-G.}\ \bibnamefont {Porquet}},
  \bibinfo {author} {\bibfnamefont {T.}~\bibnamefont {Venkova}}, \bibinfo
  {author} {\bibfnamefont {I.}~\bibnamefont {Deloncle}}, \bibinfo {author}
  {\bibfnamefont {M.}~\bibnamefont {Houry}}, \bibinfo {author} {\bibfnamefont
  {C.}~\bibnamefont {Theisen}}, \bibinfo {author} {\bibfnamefont
  {A.}~\bibnamefont {Astier}}, \bibinfo {author} {\bibfnamefont
  {A.}~\bibnamefont {Bauchet}}, \bibinfo {author} {\bibfnamefont
  {S.}~\bibnamefont {Lalkovski}}, \bibinfo {author} {\bibfnamefont
  {G.}~\bibnamefont {Barreau}}, \bibinfo {author} {\bibfnamefont
  {N.}~\bibnamefont {Buforn}}, \bibinfo {author} {\bibfnamefont
  {T.}~\bibnamefont {Doan}},  \emph {et~al.},\ }\href {\doibase
  10.1140/epja/i2002-10042-8} {\bibfield  {journal} {\bibinfo  {journal} {Eur.
  Phys. J. A}\ }\textbf {\bibinfo {volume} {15}},\ \bibinfo {pages} {315}
  (\bibinfo {year} {2002})}\BibitemShut {NoStop}%
\bibitem [{\citenamefont {Ressler}\ \emph {et~al.}(2010)\citenamefont
  {Ressler}, \citenamefont {Caggiano}, \citenamefont {Francy}, \citenamefont
  {Peplowski}, \citenamefont {Allmond}, \citenamefont {Beausang}, \citenamefont
  {Bernstein}, \citenamefont {Bleuel}, \citenamefont {Burke}, \citenamefont
  {Fallon}, \citenamefont {Hecht}, \citenamefont {Jordan}, \citenamefont
  {Lesher}, \citenamefont {McMahan}, \citenamefont {Palmer}, \citenamefont
  {Phair}, \citenamefont {Scielzo}, \citenamefont {Swearingen}, \citenamefont
  {Warren},\ and\ \citenamefont {Wiedeking}}]{re10}%
  \BibitemOpen
  \bibfield  {author} {\bibinfo {author} {\bibfnamefont {J.~J.}\ \bibnamefont
  {Ressler}}, \bibinfo {author} {\bibfnamefont {J.~A.}\ \bibnamefont
  {Caggiano}}, \bibinfo {author} {\bibfnamefont {C.~J.}\ \bibnamefont
  {Francy}}, \bibinfo {author} {\bibfnamefont {P.~N.}\ \bibnamefont
  {Peplowski}}, \bibinfo {author} {\bibfnamefont {J.~M.}\ \bibnamefont
  {Allmond}}, \bibinfo {author} {\bibfnamefont {C.~W.}\ \bibnamefont
  {Beausang}}, \bibinfo {author} {\bibfnamefont {L.~A.}\ \bibnamefont
  {Bernstein}}, \bibinfo {author} {\bibfnamefont {D.~L.}\ \bibnamefont
  {Bleuel}}, \bibinfo {author} {\bibfnamefont {J.~T.}\ \bibnamefont {Burke}},
  \bibinfo {author} {\bibfnamefont {P.}~\bibnamefont {Fallon}}, \bibinfo
  {author} {\bibfnamefont {A.~A.}\ \bibnamefont {Hecht}}, \bibinfo {author}
  {\bibfnamefont {D.~V.}\ \bibnamefont {Jordan}},  \emph {et~al.},\ }\href
  {\doibase 10.1103/PhysRevC.81.014301} {\bibfield  {journal} {\bibinfo
  {journal} {Phys. Rev. C}\ }\textbf {\bibinfo {volume} {81}},\ \bibinfo
  {pages} {014301} (\bibinfo {year} {2010})}\BibitemShut {NoStop}%
\bibitem [{\citenamefont {Scherillo}\ \emph {et~al.}(2004)\citenamefont
  {Scherillo}, \citenamefont {Genevey}, \citenamefont {Pinston}, \citenamefont
  {Covello}, \citenamefont {Faust}, \citenamefont {Gargano}, \citenamefont
  {Orlandi}, \citenamefont {Simpson}, \citenamefont {Tsekhanovich},\ and\
  \citenamefont {Warr}}]{sc04}%
  \BibitemOpen
  \bibfield  {author} {\bibinfo {author} {\bibfnamefont {A.}~\bibnamefont
  {Scherillo}}, \bibinfo {author} {\bibfnamefont {J.}~\bibnamefont {Genevey}},
  \bibinfo {author} {\bibfnamefont {J.~A.}\ \bibnamefont {Pinston}}, \bibinfo
  {author} {\bibfnamefont {A.}~\bibnamefont {Covello}}, \bibinfo {author}
  {\bibfnamefont {H.}~\bibnamefont {Faust}}, \bibinfo {author} {\bibfnamefont
  {A.}~\bibnamefont {Gargano}}, \bibinfo {author} {\bibfnamefont
  {R.}~\bibnamefont {Orlandi}}, \bibinfo {author} {\bibfnamefont {G.~S.}\
  \bibnamefont {Simpson}}, \bibinfo {author} {\bibfnamefont {I.}~\bibnamefont
  {Tsekhanovich}}, \ and\ \bibinfo {author} {\bibfnamefont {N.}~\bibnamefont
  {Warr}},\ }\href {\doibase 10.1103/PhysRevC.70.054318} {\bibfield  {journal}
  {\bibinfo  {journal} {Phys. Rev. C}\ }\textbf {\bibinfo {volume} {70}},\
  \bibinfo {pages} {054318} (\bibinfo {year} {2004})}\BibitemShut {NoStop}%
\bibitem [{\citenamefont {Babcock}\ \emph {et~al.}(2018)\citenamefont
  {Babcock}, \citenamefont {Klawitter}, \citenamefont {Leistenschneider},
  \citenamefont {Lascar}, \citenamefont {Barquest}, \citenamefont {Finlay},
  \citenamefont {Foster}, \citenamefont {Gallant}, \citenamefont {Hunt},
  \citenamefont {Kootte}, \citenamefont {Lan}, \citenamefont {Paul},
  \citenamefont {Phan}, \citenamefont {Reiter}, \citenamefont {Schultz},
  \citenamefont {Short}, \citenamefont {Andreoiu}, \citenamefont {Brodeur},
  \citenamefont {Dillmann}, \citenamefont {Gwinner}, \citenamefont
  {Kwiatkowski}, \citenamefont {Leach},\ and\ \citenamefont {Dilling}}]{ba18}%
  \BibitemOpen
  \bibfield  {author} {\bibinfo {author} {\bibfnamefont {C.}~\bibnamefont
  {Babcock}}, \bibinfo {author} {\bibfnamefont {R.}~\bibnamefont {Klawitter}},
  \bibinfo {author} {\bibfnamefont {E.}~\bibnamefont {Leistenschneider}},
  \bibinfo {author} {\bibfnamefont {D.}~\bibnamefont {Lascar}}, \bibinfo
  {author} {\bibfnamefont {B.~R.}\ \bibnamefont {Barquest}}, \bibinfo {author}
  {\bibfnamefont {A.}~\bibnamefont {Finlay}}, \bibinfo {author} {\bibfnamefont
  {M.}~\bibnamefont {Foster}}, \bibinfo {author} {\bibfnamefont {A.~T.}\
  \bibnamefont {Gallant}}, \bibinfo {author} {\bibfnamefont {P.}~\bibnamefont
  {Hunt}}, \bibinfo {author} {\bibfnamefont {B.}~\bibnamefont {Kootte}},
  \bibinfo {author} {\bibfnamefont {Y.}~\bibnamefont {Lan}}, \bibinfo {author}
  {\bibfnamefont {S.~F.}\ \bibnamefont {Paul}},  \emph {et~al.},\ }\href
  {\doibase 10.1103/PhysRevC.97.024312} {\bibfield  {journal} {\bibinfo
  {journal} {Phys. Rev. C}\ }\textbf {\bibinfo {volume} {97}},\ \bibinfo
  {pages} {024312} (\bibinfo {year} {2018})}\BibitemShut {NoStop}%
\bibitem [{\citenamefont {Gausemel}\ \emph {et~al.}(2004)\citenamefont
  {Gausemel}, \citenamefont {Fogelberg}, \citenamefont {Engeland},
  \citenamefont {Hjorth-Jensen}, \citenamefont {Hoff}, \citenamefont {Mach},
  \citenamefont {Mezilev},\ and\ \citenamefont {Omtvedt}}]{ga04}%
  \BibitemOpen
  \bibfield  {author} {\bibinfo {author} {\bibfnamefont {H.}~\bibnamefont
  {Gausemel}}, \bibinfo {author} {\bibfnamefont {B.}~\bibnamefont {Fogelberg}},
  \bibinfo {author} {\bibfnamefont {T.}~\bibnamefont {Engeland}}, \bibinfo
  {author} {\bibfnamefont {M.}~\bibnamefont {Hjorth-Jensen}}, \bibinfo {author}
  {\bibfnamefont {P.}~\bibnamefont {Hoff}}, \bibinfo {author} {\bibfnamefont
  {H.}~\bibnamefont {Mach}}, \bibinfo {author} {\bibfnamefont {K.~A.}\
  \bibnamefont {Mezilev}}, \ and\ \bibinfo {author} {\bibfnamefont {J.~P.}\
  \bibnamefont {Omtvedt}},\ }\href {\doibase 10.1103/PhysRevC.69.054307}
  {\bibfield  {journal} {\bibinfo  {journal} {Phys. Rev. C}\ }\textbf {\bibinfo
  {volume} {69}},\ \bibinfo {pages} {054307} (\bibinfo {year}
  {2004})}\BibitemShut {NoStop}%
\bibitem [{\citenamefont {Kameda}\ \emph {et~al.}(2012)\citenamefont {Kameda},
  \citenamefont {Kubo}, \citenamefont {Ohnishi}, \citenamefont {Kusaka},
  \citenamefont {Yoshida}, \citenamefont {Yoshida}, \citenamefont {Ohtake},
  \citenamefont {Fukuda}, \citenamefont {Takeda}, \citenamefont {Tanaka},
  \citenamefont {Inabe}, \citenamefont {Yanagisawa}, \citenamefont {Gono},
  \citenamefont {Watanabe}, \citenamefont {Otsu}, \citenamefont {Baba},
  \citenamefont {Ichihara}, \citenamefont {Yamaguchi}, \citenamefont {Takechi},
  \citenamefont {Nishimura}, \citenamefont {Ueno}, \citenamefont {Yoshimi},
  \citenamefont {Sakurai}, \citenamefont {Motobayashi}, \citenamefont {Nakao},
  \citenamefont {Mizoi}, \citenamefont {Matsushita}, \citenamefont {Ieki},
  \citenamefont {Kobayashi}, \citenamefont {Tanaka}, \citenamefont {Kawada},
  \citenamefont {Tanaka}, \citenamefont {Deguchi}, \citenamefont {Satou},
  \citenamefont {Kondo}, \citenamefont {Nakamura}, \citenamefont {Yoshinaga},
  \citenamefont {Ishii}, \citenamefont {Yoshii}, \citenamefont {Miyashita},
  \citenamefont {Uematsu}, \citenamefont {Shiraki}, \citenamefont {Sumikama},
  \citenamefont {Chiba}, \citenamefont {Ideguchi}, \citenamefont {Saito},
  \citenamefont {Yamaguchi}, \citenamefont {Hachiuma}, \citenamefont {Suzuki},
  \citenamefont {Moriguchi}, \citenamefont {Ozawa}, \citenamefont {Ohtsubo},
  \citenamefont {Famiano}, \citenamefont {Geissel}, \citenamefont {Nettleton},
  \citenamefont {Tarasov}, \citenamefont {Bazin}, \citenamefont {Sherrill},
  \citenamefont {Manikonda},\ and\ \citenamefont {Nolen}}]{ka12}%
  \BibitemOpen
  \bibfield  {author} {\bibinfo {author} {\bibfnamefont {D.}~\bibnamefont
  {Kameda}}, \bibinfo {author} {\bibfnamefont {T.}~\bibnamefont {Kubo}},
  \bibinfo {author} {\bibfnamefont {T.}~\bibnamefont {Ohnishi}}, \bibinfo
  {author} {\bibfnamefont {K.}~\bibnamefont {Kusaka}}, \bibinfo {author}
  {\bibfnamefont {A.}~\bibnamefont {Yoshida}}, \bibinfo {author} {\bibfnamefont
  {K.}~\bibnamefont {Yoshida}}, \bibinfo {author} {\bibfnamefont
  {M.}~\bibnamefont {Ohtake}}, \bibinfo {author} {\bibfnamefont
  {N.}~\bibnamefont {Fukuda}}, \bibinfo {author} {\bibfnamefont
  {H.}~\bibnamefont {Takeda}}, \bibinfo {author} {\bibfnamefont
  {K.}~\bibnamefont {Tanaka}}, \bibinfo {author} {\bibfnamefont
  {N.}~\bibnamefont {Inabe}}, \bibinfo {author} {\bibfnamefont
  {Y.}~\bibnamefont {Yanagisawa}},  \emph {et~al.},\ }\href {\doibase
  10.1103/PhysRevC.86.054319} {\bibfield  {journal} {\bibinfo  {journal} {Phys.
  Rev. C}\ }\textbf {\bibinfo {volume} {86}},\ \bibinfo {pages} {054319}
  (\bibinfo {year} {2012})}\BibitemShut {NoStop}%
\bibitem [{\citenamefont {Cl\'ement}\ \emph {et~al.}(2017)\citenamefont
  {Cl\'ement}, \citenamefont {Michelagnoli}, \citenamefont {de~France},
  \citenamefont {Li}, \citenamefont {Lemasson}, \citenamefont {Barthe-Dejean},
  \citenamefont {Beuzard}, \citenamefont {Bougault}, \citenamefont {Cacitti},
  \citenamefont {Foucher}, \citenamefont {Fremont}, \citenamefont {Gangnant},
  \citenamefont {Goupil}, \citenamefont {Houarner}, \citenamefont {Jean},
  \citenamefont {Lefevre}, \citenamefont {Legeard}, \citenamefont {Legruel},
  \citenamefont {Maugeais}, \citenamefont {Ménager}, \citenamefont {Ménard},
  \citenamefont {Munoz}, \citenamefont {Ozille}, \citenamefont {Raine},
  \citenamefont {Ropert}, \citenamefont {Saillant}, \citenamefont {Spitaels},
  \citenamefont {Tripon}, \citenamefont {Vallerand}, \citenamefont {Voltolini},
  \citenamefont {Korten}, \citenamefont {Salsac}, \citenamefont {Theisen},
  \citenamefont {Zielińska}, \citenamefont {Joannem}, \citenamefont {Karolak},
  \citenamefont {Kebbiri}, \citenamefont {Lotode}, \citenamefont {Touzery},
  \citenamefont {Walter}, \citenamefont {Korichi}, \citenamefont {Ljungvall},
  \citenamefont {Lopez-Martens}, \citenamefont {Ralet}, \citenamefont {Dosme},
  \citenamefont {Grave}, \citenamefont {Karkour}, \citenamefont {Lafay},
  \citenamefont {Legay}, \citenamefont {Kojouharov}, \citenamefont
  {Domingo-Pardo}, \citenamefont {Gadea}, \citenamefont {Pérez-Vidal},
  \citenamefont {Civera}, \citenamefont {Birkenbach}, \citenamefont {Eberth},
  \citenamefont {Hess}, \citenamefont {Lewandowski}, \citenamefont {Reiter},
  \citenamefont {Nannini}, \citenamefont {Angelis}, \citenamefont {Jaworski},
  \citenamefont {John}, \citenamefont {Napoli}, \citenamefont
  {Valiente-Dobón}, \citenamefont {Barrientos}, \citenamefont {Bortolato},
  \citenamefont {Benzoni}, \citenamefont {Bracco}, \citenamefont {Brambilla},
  \citenamefont {Camera}, \citenamefont {Crespi}, \citenamefont {Leoni},
  \citenamefont {Million}, \citenamefont {Pullia}, \citenamefont {Wieland},
  \citenamefont {Bazzacco}, \citenamefont {Lenzi}, \citenamefont {Lunardi},
  \citenamefont {Menegazzo}, \citenamefont {Mengoni}, \citenamefont {Recchia},
  \citenamefont {Bellato}, \citenamefont {Isocrate}, \citenamefont {Canet},
  \citenamefont {Didierjean}, \citenamefont {Duchêne}, \citenamefont
  {Baumann}, \citenamefont {Brucker}, \citenamefont {Dangelser}, \citenamefont
  {Filliger}, \citenamefont {Friedmann}, \citenamefont {Gaudiot}, \citenamefont
  {Grapton}, \citenamefont {Kocher}, \citenamefont {Mathieu}, \citenamefont
  {Sigward}, \citenamefont {Thomas}, \citenamefont {Veeramootoo}, \citenamefont
  {Dudouet}, \citenamefont {Stézowski}, \citenamefont {Aufranc}, \citenamefont
  {Aubert}, \citenamefont {Labiche}, \citenamefont {Simpson}, \citenamefont
  {Burrows}, \citenamefont {Coleman-Smith}, \citenamefont {Grant},
  \citenamefont {Lazarus}, \citenamefont {Morrall}, \citenamefont {Pucknell},
  \citenamefont {Boston}, \citenamefont {Judson}, \citenamefont {Lalović},
  \citenamefont {Nyberg}, \citenamefont {Collado}, \citenamefont {González},
  \citenamefont {Kuti}, \citenamefont {Nyakó}, \citenamefont {Maj},\ and\
  \citenamefont {Rudigier}}]{cl16}%
  \BibitemOpen
  \bibfield  {author} {\bibinfo {author} {\bibfnamefont {E.}~\bibnamefont
  {Cl\'ement}}, \bibinfo {author} {\bibfnamefont {C.}~\bibnamefont
  {Michelagnoli}}, \bibinfo {author} {\bibfnamefont {G.}~\bibnamefont
  {de~France}}, \bibinfo {author} {\bibfnamefont {H.}~\bibnamefont {Li}},
  \bibinfo {author} {\bibfnamefont {A.}~\bibnamefont {Lemasson}}, \bibinfo
  {author} {\bibfnamefont {C.}~\bibnamefont {Barthe-Dejean}}, \bibinfo {author}
  {\bibfnamefont {M.}~\bibnamefont {Beuzard}}, \bibinfo {author} {\bibfnamefont
  {P.}~\bibnamefont {Bougault}}, \bibinfo {author} {\bibfnamefont
  {J.}~\bibnamefont {Cacitti}}, \bibinfo {author} {\bibfnamefont {J.-L.}\
  \bibnamefont {Foucher}}, \bibinfo {author} {\bibfnamefont {G.}~\bibnamefont
  {Fremont}}, \bibinfo {author} {\bibfnamefont {P.}~\bibnamefont {Gangnant}},
  \emph {et~al.},\ }\href {\doibase 10.1016/j.nima.2017.02.063} {\bibfield
  {journal} {\bibinfo  {journal} {Nucl. Instr. and Meth. A}\ }\textbf {\bibinfo
  {volume} {855}},\ \bibinfo {pages} {1 } (\bibinfo {year} {2017})}\BibitemShut
  {NoStop}%
\bibitem [{\citenamefont {Akkoyun}\ \emph {et~al.}(2012)\citenamefont
  {Akkoyun}, \citenamefont {Algora}, \citenamefont {Alikhani}, \citenamefont
  {Ameil}, \citenamefont {de~Angelis}, \citenamefont {Arnold}, \citenamefont
  {Astier}, \citenamefont {Ataç}, \citenamefont {Aubert}, \citenamefont
  {Aufranc}, \citenamefont {Austin}, \citenamefont {Aydin}, \citenamefont
  {Azaiez}, \citenamefont {Badoer}, \citenamefont {Balabanski}, \citenamefont
  {Barrientos}, \citenamefont {Baulieu}, \citenamefont {Baumann}, \citenamefont
  {Bazzacco}, \citenamefont {Beck}, \citenamefont {Beck}, \citenamefont
  {Bednarczyk}, \citenamefont {Bellato}, \citenamefont {Bentley}, \citenamefont
  {Benzoni}, \citenamefont {Berthier}, \citenamefont {Berti}, \citenamefont
  {Beunard}, \citenamefont {Bianco}, \citenamefont {Birkenbach}, \citenamefont
  {Bizzeti}, \citenamefont {Bizzeti-Sona}, \citenamefont {Blanc}, \citenamefont
  {Blasco}, \citenamefont {Blasi}, \citenamefont {Bloor}, \citenamefont
  {Boiano}, \citenamefont {Borsato}, \citenamefont {Bortolato}, \citenamefont
  {Boston}, \citenamefont {Boston}, \citenamefont {Bourgault}, \citenamefont
  {Boutachkov}, \citenamefont {Bouty}, \citenamefont {Bracco}, \citenamefont
  {Brambilla}, \citenamefont {Brawn}, \citenamefont {Brondi}, \citenamefont
  {Broussard}, \citenamefont {Bruyneel}, \citenamefont {Bucurescu},
  \citenamefont {Burrows}, \citenamefont {Bürger}, \citenamefont {Cabaret},
  \citenamefont {Cahan}, \citenamefont {Calore}, \citenamefont {Camera},
  \citenamefont {Capsoni}, \citenamefont {Carrió}, \citenamefont {Casati},
  \citenamefont {Castoldi}, \citenamefont {Cederwall}, \citenamefont {Cercus},
  \citenamefont {Chambert}, \citenamefont {Chambit}, \citenamefont {Chapman},
  \citenamefont {Charles}, \citenamefont {Chavas}, \citenamefont {Clément},
  \citenamefont {Cocconi}, \citenamefont {Coelli}, \citenamefont
  {Coleman-Smith}, \citenamefont {Colombo}, \citenamefont {Colosimo},
  \citenamefont {Commeaux}, \citenamefont {Conventi}, \citenamefont {Cooper},
  \citenamefont {Corsi}, \citenamefont {Cortesi}, \citenamefont {Costa},
  \citenamefont {Crespi}, \citenamefont {Cresswell}, \citenamefont {Cullen},
  \citenamefont {Curien}, \citenamefont {Czermak}, \citenamefont {Delbourg},
  \citenamefont {Depalo}, \citenamefont {Descombes}, \citenamefont
  {Désesquelles}, \citenamefont {Detistov}, \citenamefont {Diarra},
  \citenamefont {Didierjean}, \citenamefont {Dimmock}, \citenamefont {Doan},
  \citenamefont {Domingo-Pardo}, \citenamefont {Doncel}, \citenamefont
  {Dorangeville}, \citenamefont {Dosme}, \citenamefont {Drouen}, \citenamefont
  {Duchêne}, \citenamefont {Dulny}, \citenamefont {Eberth}, \citenamefont
  {Edelbruck}, \citenamefont {Egea}, \citenamefont {Engert}, \citenamefont
  {Erduran}, \citenamefont {Ertürk}, \citenamefont {Fanin}, \citenamefont
  {Fantinel}, \citenamefont {Farnea}, \citenamefont {Faul}, \citenamefont
  {Filliger}, \citenamefont {Filmer}, \citenamefont {Finck}, \citenamefont
  {de~France}, \citenamefont {Gadea}, \citenamefont {Gast}, \citenamefont
  {Geraci}, \citenamefont {Gerl}, \citenamefont {Gernhäuser}, \citenamefont
  {Giannatiempo}, \citenamefont {Giaz}, \citenamefont {Gibelin}, \citenamefont
  {Givechev}, \citenamefont {Goel}, \citenamefont {González}, \citenamefont
  {Gottardo}, \citenamefont {Grave}, \citenamefont {Gre¸bosz}, \citenamefont
  {Griffiths}, \citenamefont {Grint}, \citenamefont {Gros}, \citenamefont
  {Guevara}, \citenamefont {Gulmini}, \citenamefont {Görgen}, \citenamefont
  {Ha}, \citenamefont {Habermann}, \citenamefont {Harkness}, \citenamefont
  {Harroch}, \citenamefont {Hauschild}, \citenamefont {He}, \citenamefont
  {Hernández-Prieto}, \citenamefont {Hervieu}, \citenamefont {Hess},
  \citenamefont {Hüyük}, \citenamefont {Ince}, \citenamefont {Isocrate},
  \citenamefont {Jaworski}, \citenamefont {Johnson}, \citenamefont {Jolie},
  \citenamefont {Jones}, \citenamefont {Jonson}, \citenamefont {Joshi},
  \citenamefont {Judson}, \citenamefont {Jungclaus}, \citenamefont {Kaci},
  \citenamefont {Karkour}, \citenamefont {Karolak}, \citenamefont {Kaşkaş},
  \citenamefont {Kebbiri}, \citenamefont {Kempley}, \citenamefont {Khaplanov},
  \citenamefont {Klupp}, \citenamefont {Kogimtzis}, \citenamefont {Kojouharov},
  \citenamefont {Korichi}, \citenamefont {Korten}, \citenamefont {Kröll},
  \citenamefont {Krücken}, \citenamefont {Kurz}, \citenamefont {Ky},
  \citenamefont {Labiche}, \citenamefont {Lafay}, \citenamefont {Lavergne},
  \citenamefont {Lazarus}, \citenamefont {Leboutelier}, \citenamefont
  {Lefebvre}, \citenamefont {Legay}, \citenamefont {Legeard}, \citenamefont
  {Lelli}, \citenamefont {Lenzi}, \citenamefont {Leoni}, \citenamefont
  {Lermitage}, \citenamefont {Lersch}, \citenamefont {Leske}, \citenamefont
  {Letts}, \citenamefont {Lhenoret}, \citenamefont {Lieder}, \citenamefont
  {Linget}, \citenamefont {Ljungvall}, \citenamefont {Lopez-Martens},
  \citenamefont {Lotodé}, \citenamefont {Lunardi}, \citenamefont {Maj},
  \citenamefont {van~der Marel}, \citenamefont {Mariette}, \citenamefont
  {Marginean}, \citenamefont {Marginean}, \citenamefont {Maron}, \citenamefont
  {Mather}, \citenamefont {Me¸czyński}, \citenamefont {Mendéz},
  \citenamefont {Medina}, \citenamefont {Melon}, \citenamefont {Menegazzo},
  \citenamefont {Mengoni}, \citenamefont {Merchan}, \citenamefont {Mihailescu},
  \citenamefont {Michelagnoli}, \citenamefont {Mierzejewski}, \citenamefont
  {Milechina}, \citenamefont {Million}, \citenamefont {Mitev}, \citenamefont
  {Molini}, \citenamefont {Montanari}, \citenamefont {Moon}, \citenamefont
  {Morbiducci}, \citenamefont {Moro}, \citenamefont {Morrall}, \citenamefont
  {Möller}, \citenamefont {Nannini}, \citenamefont {Napoli}, \citenamefont
  {Nelson}, \citenamefont {Nespolo}, \citenamefont {Ngo}, \citenamefont
  {Nicoletto}, \citenamefont {Nicolini}, \citenamefont {Noa}, \citenamefont
  {Nolan}, \citenamefont {Norman}, \citenamefont {Nyberg}, \citenamefont
  {Obertelli}, \citenamefont {Olariu}, \citenamefont {Orlandi}, \citenamefont
  {Oxley}, \citenamefont {Özben}, \citenamefont {Ozille}, \citenamefont
  {Oziol}, \citenamefont {Pachoud}, \citenamefont {Palacz}, \citenamefont
  {Palin}, \citenamefont {Pancin}, \citenamefont {Parisel}, \citenamefont
  {Pariset}, \citenamefont {Pascovici}, \citenamefont {Peghin}, \citenamefont
  {Pellegri}, \citenamefont {Perego}, \citenamefont {Perrier}, \citenamefont
  {Petcu}, \citenamefont {Petkov}, \citenamefont {Petrache}, \citenamefont
  {Pierre}, \citenamefont {Pietralla}, \citenamefont {Pietri}, \citenamefont
  {Pignanelli}, \citenamefont {Piqueras}, \citenamefont {Podolyak},
  \citenamefont {Pouhalec}, \citenamefont {Pouthas}, \citenamefont {Pugnére},
  \citenamefont {Pucknell}, \citenamefont {Pullia}, \citenamefont {Quintana},
  \citenamefont {Raine}, \citenamefont {Rainovski}, \citenamefont {Ramina},
  \citenamefont {Rampazzo}, \citenamefont {Rana}, \citenamefont {Rebeschini},
  \citenamefont {Recchia}, \citenamefont {Redon}, \citenamefont {Reese},
  \citenamefont {Reiter}, \citenamefont {Regan}, \citenamefont {Riboldi},
  \citenamefont {Richer}, \citenamefont {Rigato}, \citenamefont {Rigby},
  \citenamefont {Ripamonti}, \citenamefont {Robinson}, \citenamefont {Robin},
  \citenamefont {Roccaz}, \citenamefont {Ropert}, \citenamefont {Rossé},
  \citenamefont {Alvarez}, \citenamefont {Rosso}, \citenamefont {Rubio},
  \citenamefont {Rudolph}, \citenamefont {Saillant}, \citenamefont {Şahin},
  \citenamefont {Salomon}, \citenamefont {Salsac}, \citenamefont {Salt},
  \citenamefont {Salvato}, \citenamefont {Sampson}, \citenamefont {Sanchis},
  \citenamefont {Santos}, \citenamefont {Schaffner}, \citenamefont {Schlarb},
  \citenamefont {Scraggs}, \citenamefont {Seddon}, \citenamefont {Şenyiğit},
  \citenamefont {Sigward}, \citenamefont {Simpson}, \citenamefont {Simpson},
  \citenamefont {Slee}, \citenamefont {Smith}, \citenamefont {Sona},
  \citenamefont {Sowicki}, \citenamefont {Spolaore}, \citenamefont {Stahl},
  \citenamefont {Stanios}, \citenamefont {Stefanova}, \citenamefont
  {Stézowski}, \citenamefont {Strachan}, \citenamefont {Suliman},
  \citenamefont {Söderström}, \citenamefont {Tain}, \citenamefont {Tanguy},
  \citenamefont {Tashenov}, \citenamefont {Theisen}, \citenamefont {Thornhill},
  \citenamefont {Tomasi}, \citenamefont {Toniolo}, \citenamefont {Touzery},
  \citenamefont {Travers}, \citenamefont {Triossi}, \citenamefont {Tripon},
  \citenamefont {Tun-Lanoë}, \citenamefont {Turcato}, \citenamefont
  {Unsworth}, \citenamefont {Ur}, \citenamefont {Valiente-Dobon}, \citenamefont
  {Vandone}, \citenamefont {Vardaci}, \citenamefont {Venturelli}, \citenamefont
  {Veronese}, \citenamefont {Veyssiere}, \citenamefont {Viscione},
  \citenamefont {Wadsworth}, \citenamefont {Walker}, \citenamefont {Warr},
  \citenamefont {Weber}, \citenamefont {Weisshaar}, \citenamefont {Wells},
  \citenamefont {Wieland}, \citenamefont {Wiens}, \citenamefont {Wittwer},
  \citenamefont {Wollersheim}, \citenamefont {Zocca}, \citenamefont {Zamfir},
  \citenamefont {Zie¸bliński},\ and\ \citenamefont {Zucchiatti}}]{ak12}%
  \BibitemOpen
  \bibfield  {author} {\bibinfo {author} {\bibfnamefont {S.}~\bibnamefont
  {Akkoyun}}, \bibinfo {author} {\bibfnamefont {A.}~\bibnamefont {Algora}},
  \bibinfo {author} {\bibfnamefont {B.}~\bibnamefont {Alikhani}}, \bibinfo
  {author} {\bibfnamefont {F.}~\bibnamefont {Ameil}}, \bibinfo {author}
  {\bibfnamefont {G.}~\bibnamefont {de~Angelis}}, \bibinfo {author}
  {\bibfnamefont {L.}~\bibnamefont {Arnold}}, \bibinfo {author} {\bibfnamefont
  {A.}~\bibnamefont {Astier}}, \bibinfo {author} {\bibfnamefont
  {A.}~\bibnamefont {Ataç}}, \bibinfo {author} {\bibfnamefont
  {Y.}~\bibnamefont {Aubert}}, \bibinfo {author} {\bibfnamefont
  {C.}~\bibnamefont {Aufranc}}, \bibinfo {author} {\bibfnamefont
  {A.}~\bibnamefont {Austin}}, \bibinfo {author} {\bibfnamefont
  {S.}~\bibnamefont {Aydin}},  \emph {et~al.},\ }\href {\doibase
  10.1016/j.nima.2011.11.081} {\bibfield  {journal} {\bibinfo  {journal} {Nucl.
  Instr. and Meth. A}\ }\textbf {\bibinfo {volume} {668}},\ \bibinfo {pages}
  {26 } (\bibinfo {year} {2012})}\BibitemShut {NoStop}%
\bibitem [{\citenamefont {Rejmund}\ \emph {et~al.}(2011)\citenamefont
  {Rejmund}, \citenamefont {Lecornu}, \citenamefont {Navin}, \citenamefont
  {Schmitt}, \citenamefont {Damoy}, \citenamefont {Delaune}, \citenamefont
  {Enguerrand}, \citenamefont {Fremont}, \citenamefont {Gangnant},
  \citenamefont {Gaudefroy}, \citenamefont {Jacquot}, \citenamefont {Pancin},
  \citenamefont {Pullanhiotan},\ and\ \citenamefont {Spitaels}}]{re11}%
  \BibitemOpen
  \bibfield  {author} {\bibinfo {author} {\bibfnamefont {M.}~\bibnamefont
  {Rejmund}}, \bibinfo {author} {\bibfnamefont {B.}~\bibnamefont {Lecornu}},
  \bibinfo {author} {\bibfnamefont {A.}~\bibnamefont {Navin}}, \bibinfo
  {author} {\bibfnamefont {C.}~\bibnamefont {Schmitt}}, \bibinfo {author}
  {\bibfnamefont {S.}~\bibnamefont {Damoy}}, \bibinfo {author} {\bibfnamefont
  {O.}~\bibnamefont {Delaune}}, \bibinfo {author} {\bibfnamefont {J.~M.}\
  \bibnamefont {Enguerrand}}, \bibinfo {author} {\bibfnamefont
  {G.}~\bibnamefont {Fremont}}, \bibinfo {author} {\bibfnamefont
  {P.}~\bibnamefont {Gangnant}}, \bibinfo {author} {\bibfnamefont
  {L.}~\bibnamefont {Gaudefroy}}, \bibinfo {author} {\bibfnamefont
  {B.}~\bibnamefont {Jacquot}}, \bibinfo {author} {\bibfnamefont
  {J.}~\bibnamefont {Pancin}},  \emph {et~al.},\ }\href {\doibase
  10.1016/j.nima.2011.05.007} {\bibfield  {journal} {\bibinfo  {journal} {Nucl.
  Instr. and Meth. A}\ }\textbf {\bibinfo {volume} {646}},\ \bibinfo {pages}
  {184} (\bibinfo {year} {2011})}\BibitemShut {NoStop}%
\bibitem [{\citenamefont {Simpson}\ \emph {et~al.}(2000)\citenamefont
  {Simpson}, \citenamefont {Azaiez}, \citenamefont {DeFrance}, \citenamefont
  {Fouan}, \citenamefont {Gerl}, \citenamefont {Julin}, \citenamefont {Korten},
  \citenamefont {Nolan}, \citenamefont {Nyak{\'o}}, \citenamefont {Sletten},\
  and\ \citenamefont {Walker}}]{si00}%
  \BibitemOpen
  \bibfield  {author} {\bibinfo {author} {\bibfnamefont {J.}~\bibnamefont
  {Simpson}}, \bibinfo {author} {\bibfnamefont {F.}~\bibnamefont {Azaiez}},
  \bibinfo {author} {\bibfnamefont {G.}~\bibnamefont {DeFrance}}, \bibinfo
  {author} {\bibfnamefont {J.}~\bibnamefont {Fouan}}, \bibinfo {author}
  {\bibfnamefont {J.}~\bibnamefont {Gerl}}, \bibinfo {author} {\bibfnamefont
  {R.}~\bibnamefont {Julin}}, \bibinfo {author} {\bibfnamefont
  {W.}~\bibnamefont {Korten}}, \bibinfo {author} {\bibfnamefont
  {P.}~\bibnamefont {Nolan}}, \bibinfo {author} {\bibfnamefont
  {B.}~\bibnamefont {Nyak{\'o}}}, \bibinfo {author} {\bibfnamefont
  {G.}~\bibnamefont {Sletten}}, \ and\ \bibinfo {author} {\bibfnamefont
  {P.}~\bibnamefont {Walker}},\ }\href@noop {} {\bibfield  {journal} {\bibinfo
  {journal} {Act. Phys. Hung. Ser. Heavy Ion Phys.}\ }\textbf {\bibinfo
  {volume} {11}},\ \bibinfo {pages} {159} (\bibinfo {year} {2000})}\BibitemShut
  {NoStop}%
\bibitem [{\citenamefont {Navin}\ \emph {et~al.}(2014)\citenamefont {Navin},
  \citenamefont {Rejmund}, \citenamefont {Schmitt}, \citenamefont
  {Bhattacharyya}, \citenamefont {Lhersonneau}, \citenamefont {van Isacker},
  \citenamefont {Caama{\~{n}}o}, \citenamefont {Cl{\'{e}}ment}, \citenamefont
  {Delaune}, \citenamefont {Farget}, \citenamefont {de~France},\ and\
  \citenamefont {Jacquot}}]{na14}%
  \BibitemOpen
  \bibfield  {author} {\bibinfo {author} {\bibfnamefont {A.}~\bibnamefont
  {Navin}}, \bibinfo {author} {\bibfnamefont {M.}~\bibnamefont {Rejmund}},
  \bibinfo {author} {\bibfnamefont {C.}~\bibnamefont {Schmitt}}, \bibinfo
  {author} {\bibfnamefont {S.}~\bibnamefont {Bhattacharyya}}, \bibinfo {author}
  {\bibfnamefont {G.}~\bibnamefont {Lhersonneau}}, \bibinfo {author}
  {\bibfnamefont {P.}~\bibnamefont {van Isacker}}, \bibinfo {author}
  {\bibfnamefont {M.}~\bibnamefont {Caama{\~{n}}o}}, \bibinfo {author}
  {\bibfnamefont {E.}~\bibnamefont {Cl{\'{e}}ment}}, \bibinfo {author}
  {\bibfnamefont {O.}~\bibnamefont {Delaune}}, \bibinfo {author} {\bibfnamefont
  {F.}~\bibnamefont {Farget}}, \bibinfo {author} {\bibfnamefont
  {G.}~\bibnamefont {de~France}}, \ and\ \bibinfo {author} {\bibfnamefont
  {B.}~\bibnamefont {Jacquot}},\ }\href {\doibase
  10.1016/j.physletb.2013.11.024} {\bibfield  {journal} {\bibinfo  {journal}
  {Phys. Lett. B}\ }\textbf {\bibinfo {volume} {728}},\ \bibinfo {pages} {136}
  (\bibinfo {year} {2014})}\BibitemShut {NoStop}%
\bibitem [{\citenamefont {Navin}(2014)\citenamefont {Navin}\ and\ \citenamefont
  {Rejmund}}]{nare}%
  \BibitemOpen
  \bibfield  {author} {\bibinfo {author} {\bibfnamefont {A.}~\bibnamefont
  {Navin}}\ and\ \bibinfo {author} {\bibfnamefont {M.}~\bibnamefont
  {Rejmund}},\ }\enquote {\bibinfo {title} {{Gamma-ray spectroscopy of
  neutron-rich fission fragments}},}\ in\ \href {\doibase
  10.1036/1097-8542.YB140316} {\emph {\bibinfo {booktitle} {Yearkbook of
  Encyclopedia of Science and Technology}}}\ (\bibinfo  {publisher}
  {McGraw-Hill},\ \bibinfo {year} {2014})\BibitemShut {NoStop}%
\bibitem [{\citenamefont {Kim}\ \emph {et~al.}(2017)\citenamefont {Kim},
  \citenamefont {Lemasson}, \citenamefont {Rejmund}, \citenamefont {Navin},
  \citenamefont {Biswas}, \citenamefont {Michelagnoli}, \citenamefont {Stefan},
  \citenamefont {Banik}, \citenamefont {Bednarczyk}, \citenamefont
  {Bhattacharya}, \citenamefont {Bhattacharyya}, \citenamefont {Cl{\'{e}}ment},
  \citenamefont {Crawford}, \citenamefont {{De France}}, \citenamefont
  {Fallon}, \citenamefont {Goupil}, \citenamefont {Jacquot}, \citenamefont
  {Li}, \citenamefont {Ljungvall}, \citenamefont {Macchiavelli}, \citenamefont
  {Maj}, \citenamefont {M{\'{e}}nager}, \citenamefont {Morel}, \citenamefont
  {Palit}, \citenamefont {P{\'{e}}rez-Vidal}, \citenamefont {Ropert},\ and\
  \citenamefont {Schmitt}}]{ki17}%
  \BibitemOpen
  \bibfield  {author} {\bibinfo {author} {\bibfnamefont {Y.~H.}\ \bibnamefont
  {Kim}}, \bibinfo {author} {\bibfnamefont {A.}~\bibnamefont {Lemasson}},
  \bibinfo {author} {\bibfnamefont {M.}~\bibnamefont {Rejmund}}, \bibinfo
  {author} {\bibfnamefont {A.}~\bibnamefont {Navin}}, \bibinfo {author}
  {\bibfnamefont {S.}~\bibnamefont {Biswas}}, \bibinfo {author} {\bibfnamefont
  {C.}~\bibnamefont {Michelagnoli}}, \bibinfo {author} {\bibfnamefont
  {I.}~\bibnamefont {Stefan}}, \bibinfo {author} {\bibfnamefont
  {R.}~\bibnamefont {Banik}}, \bibinfo {author} {\bibfnamefont
  {P.}~\bibnamefont {Bednarczyk}}, \bibinfo {author} {\bibfnamefont
  {S.}~\bibnamefont {Bhattacharya}}, \bibinfo {author} {\bibfnamefont
  {S.}~\bibnamefont {Bhattacharyya}}, \bibinfo {author} {\bibfnamefont
  {E.}~\bibnamefont {Cl{\'{e}}ment}},  \emph {et~al.},\ }\href {\doibase
  10.1140/epja/i2017-12353-y} {\bibfield  {journal} {\bibinfo  {journal} {Eur.
  Phys. J. A}\ }\textbf {\bibinfo {volume} {53}},\ \bibinfo {pages} {162}
  (\bibinfo {year} {2017})}\BibitemShut {NoStop}%
\bibitem [{\citenamefont {Vandebrouck}\ \emph {et~al.}(2016)\citenamefont
  {Vandebrouck}, \citenamefont {Lemasson}, \citenamefont {Rejmund},
  \citenamefont {Fremont}, \citenamefont {Pancin}, \citenamefont {Navin},
  \citenamefont {Michelagnoli}, \citenamefont {Goupil}, \citenamefont
  {Spitaels},\ and\ \citenamefont {Jacquot}}]{va16}%
  \BibitemOpen
  \bibfield  {author} {\bibinfo {author} {\bibfnamefont {M.}~\bibnamefont
  {Vandebrouck}}, \bibinfo {author} {\bibfnamefont {A.}~\bibnamefont
  {Lemasson}}, \bibinfo {author} {\bibfnamefont {M.}~\bibnamefont {Rejmund}},
  \bibinfo {author} {\bibfnamefont {G.}~\bibnamefont {Fremont}}, \bibinfo
  {author} {\bibfnamefont {J.}~\bibnamefont {Pancin}}, \bibinfo {author}
  {\bibfnamefont {A.}~\bibnamefont {Navin}}, \bibinfo {author} {\bibfnamefont
  {C.}~\bibnamefont {Michelagnoli}}, \bibinfo {author} {\bibfnamefont
  {J.}~\bibnamefont {Goupil}}, \bibinfo {author} {\bibfnamefont
  {C.}~\bibnamefont {Spitaels}}, \ and\ \bibinfo {author} {\bibfnamefont
  {B.}~\bibnamefont {Jacquot}},\ }\href {\doibase 10.1016/j.nima.2015.12.040}
  {\bibfield  {journal} {\bibinfo  {journal} {Nucl. Instr. and Meth. A}\
  }\textbf {\bibinfo {volume} {812}},\ \bibinfo {pages} {112} (\bibinfo {year}
  {2016})}\BibitemShut {NoStop}%
\bibitem [{\citenamefont {Biswas}\ \emph {et~al.}(2019)\citenamefont {Biswas},
  \citenamefont {Lemasson}, \citenamefont {Rejmund}, \citenamefont {Navin},
  \citenamefont {Kim}, \citenamefont {Michelagnoli}, \citenamefont {Stefan},
  \citenamefont {Banik}, \citenamefont {Bednarczyk}, \citenamefont
  {Bhattacharya}, \citenamefont {Bhattacharyya}, \citenamefont {Cl\'ement},
  \citenamefont {Crawford}, \citenamefont {de~France}, \citenamefont {Fallon},
  \citenamefont {Fr\'emont}, \citenamefont {Goupil}, \citenamefont {Jacquot},
  \citenamefont {Li}, \citenamefont {Ljungvall}, \citenamefont {Maj},
  \citenamefont {M\'enager}, \citenamefont {Morel}, \citenamefont {Palit},
  \citenamefont {P\'erez-Vidal}, \citenamefont {Ropert}, \citenamefont
  {Barrientos}, \citenamefont {Benzoni}, \citenamefont {Birkenbach},
  \citenamefont {Boston}, \citenamefont {Boston}, \citenamefont {Cederwall},
  \citenamefont {Collado}, \citenamefont {Cullen}, \citenamefont
  {D\'esesquelles}, \citenamefont {Domingo-Pardo}, \citenamefont {Dudouet},
  \citenamefont {Eberth}, \citenamefont {Gonz\'alez}, \citenamefont
  {Harkness-Brennan}, \citenamefont {Hess}, \citenamefont {Jungclaus},
  \citenamefont {Korten}, \citenamefont {Labiche}, \citenamefont {Lefevre},
  \citenamefont {Menegazzo}, \citenamefont {Mengoni}, \citenamefont {Million},
  \citenamefont {Napoli}, \citenamefont {Pullia}, \citenamefont {Quintana},
  \citenamefont {Ralet}, \citenamefont {Recchia}, \citenamefont {Reiter},
  \citenamefont {Saillant}, \citenamefont {Salsac}, \citenamefont {Sanchis},
  \citenamefont {Stezowski}, \citenamefont {Theisen}, \citenamefont
  {Valiente-Dob\'on},\ and\ \citenamefont {Zieli\ifmmode~\acute{n}\else
  \'{n}\fi{}ska}}]{sb19}%
  \BibitemOpen
  \bibfield  {author} {\bibinfo {author} {\bibfnamefont {S.}~\bibnamefont
  {Biswas}}, \bibinfo {author} {\bibfnamefont {A.}~\bibnamefont {Lemasson}},
  \bibinfo {author} {\bibfnamefont {M.}~\bibnamefont {Rejmund}}, \bibinfo
  {author} {\bibfnamefont {A.}~\bibnamefont {Navin}}, \bibinfo {author}
  {\bibfnamefont {Y.~H.}\ \bibnamefont {Kim}}, \bibinfo {author} {\bibfnamefont
  {C.}~\bibnamefont {Michelagnoli}}, \bibinfo {author} {\bibfnamefont
  {I.}~\bibnamefont {Stefan}}, \bibinfo {author} {\bibfnamefont
  {R.}~\bibnamefont {Banik}}, \bibinfo {author} {\bibfnamefont
  {P.}~\bibnamefont {Bednarczyk}}, \bibinfo {author} {\bibfnamefont
  {S.}~\bibnamefont {Bhattacharya}}, \bibinfo {author} {\bibfnamefont
  {S.}~\bibnamefont {Bhattacharyya}}, \bibinfo {author} {\bibfnamefont
  {E.}~\bibnamefont {Cl\'ement}},  \emph {et~al.},\ }\href {\doibase
  10.1103/PhysRevC.99.064302} {\bibfield  {journal} {\bibinfo  {journal} {Phys.
  Rev. C}\ }\textbf {\bibinfo {volume} {99}},\ \bibinfo {pages} {064302}
  (\bibinfo {year} {2019})}\BibitemShut {NoStop}%
\bibitem [{Sup(2020)}]{Sup}%
  \BibitemOpen
  \href@noop {} {\enquote {\bibinfo {title} {See the supplemental material at
  [url] for the tabulated energies and intensites of the $\gamma$-ray
  transitions observed in this work (tables {S1} and {S2})},}\ } (\bibinfo
  {year} {2020})\BibitemShut {NoStop}%
\bibitem [{\citenamefont {Hjorth-Jensen}\ \emph {et~al.}(1995)\citenamefont
  {Hjorth-Jensen}, \citenamefont {Kuo},\ and\ \citenamefont {Osnes}}]{hj95}%
  \BibitemOpen
  \bibfield  {author} {\bibinfo {author} {\bibfnamefont {M.}~\bibnamefont
  {Hjorth-Jensen}}, \bibinfo {author} {\bibfnamefont {T.~T.}\ \bibnamefont
  {Kuo}}\ and\ \bibinfo {author} {\bibfnamefont {E.}~\bibnamefont {Osnes}},\
  }\href {\doibase 10.1016/0370-1573(95)00012-6} {\bibfield  {journal}
  {\bibinfo  {journal} {Physics Reports}\ }\textbf {\bibinfo {volume} {261}},\
  \bibinfo {pages} {125 } (\bibinfo {year} {1995})}\BibitemShut {NoStop}%
\bibitem [{\citenamefont {Caurier}\ \emph {et~al.}(1999)\citenamefont
  {Caurier}, \citenamefont {Martinez-Pinedo}, \citenamefont {Nowacki},
  \citenamefont {Poves}, \citenamefont {Retamosa},\ and\ \citenamefont
  {Zuker}}]{ca99}%
  \BibitemOpen
  \bibfield  {author} {\bibinfo {author} {\bibfnamefont {E.}~\bibnamefont
  {Caurier}}, \bibinfo {author} {\bibfnamefont {G.}~\bibnamefont
  {Martinez-Pinedo}}, \bibinfo {author} {\bibfnamefont {F.}~\bibnamefont
  {Nowacki}}, \bibinfo {author} {\bibfnamefont {A.}~\bibnamefont {Poves}},
  \bibinfo {author} {\bibfnamefont {J.}~\bibnamefont {Retamosa}}, \ and\
  \bibinfo {author} {\bibfnamefont {A.~P.}\ \bibnamefont {Zuker}},\ }\href
  {\doibase 10.1103/PhysRevC.59.2033} {\bibfield  {journal} {\bibinfo
  {journal} {Phys. Rev. C}\ }\textbf {\bibinfo {volume} {59}},\ \bibinfo
  {pages} {2033} (\bibinfo {year} {1999})}\BibitemShut {NoStop}%
\end{thebibliography}%
 \bibliographystyle{myapsrev4-1}

\end{document}